\documentclass{article}
\usepackage[utf8]{inputenc}
\usepackage{geometry}
\usepackage{physics}
\usepackage{dsfont}
\usepackage{amsmath}
\usepackage{amssymb}
\usepackage{graphicx}
\usepackage{babel,blindtext}
\usepackage{xcolor}
\usepackage{hyperref}
\usepackage{cite}
\usepackage{wrapfig}
\usepackage{caption}
\usepackage{subcaption}
\usepackage{bbm}
\usepackage{authblk}
\title{\bf Quantum correlations in a mixed spin-(1/2,1) Heisenberg dimer }

\date{}
\author[1]{\small S. Bhuvaneswari}
\author[2]{\small R. Muthuganesan}
\author[3]{\small R. Radha \footnote{Corresponding author}}
\affil[1]{\footnotesize Centre for Nonlinear Science  (CeNSc), PG \& Research Department of Physics, Government College for Women (Autonomous), Kumbakonam, Tamil Nadu, India.}
\affil[2]{\footnotesize Department of Physics, Faculty of Nuclear Sciences and Physical Engineering, Czech Technical University in Prague,
B\u rehov\'a 7, 115 19 Praha 1-Star\'e M\u{e}sto, Czech Republic, Email: rajendramuthu@gmail.com}

\begin{document}

\maketitle
\begin{abstract}
 In this article, we consider the  heterodinuclear complex [Ni(dpt)(H2 O)Cu(pba)] · 2H2 O [pba =1,3-propylenebis(oxamato) and dpt = bis-(3-aminopropyl)amine]  realized through the theoretical model of mixed spin-(1/2,1) coupled via Heisenberg interaction. We study the behaviors of thermal quantum correlations of the above material via Measurement-Induced Nonlocality (MIN) based on Hilbert-Schmidt norm and fidelity. We observe that the quantum correlation measures increase with the magnetic field in an unconventional way. The role of system parameters is also brought out at thermal equilibrium. The highlight of the results is that  we are able to show the existence of room temperature quantum correlation using fidelity based MIN whereas  the entanglement ceases to exist at $141K$. 
\end{abstract}

\vspace{1cm}
~~~~~~~~~\noindent{\it Keywords}: Quantum correlation, Heisenberg interaction, Measurements, Nonlocality.
\newpage
 \renewcommand{\thefootnote}{*}
\section{Introduction}
In an era of information and communication technology, the necessity to obtain a potential solution for classically intractable problems can be attained by performing quantum computing. The incorporation of quantum principles such as superposition, entanglement, and interference into computing makes quantum computers more efficient and effective than classical counterparts. Also, this brings in variation to our conventional notion of the principle of locality as proposed in EPR paradox \cite{Schor,Eins,Bell,Bell12}. The nonlocality that arises due to superposition principle and entanglement is the most peculiar manifestation of nonlocality which has no classical analogy. The presence of entanglement or nonlocal character of quantum system is demonstrated by the violation of Bell inequality \cite{Bell,Bell12}. The study of entanglement in bipartite/ multipartite states has been investigated effectively for many decades and proven  that the entanglement is not the whole indicator of the non-classical correlation of quantum system even in the bipartite scenario.   In this connection, the seminal work of Werner \cite{Werner1989} and the subsequent experimental demonstration of quantum advantages using separable states \cite{WangDQC} suggest that the entanglement is not an ultimate resource for quantum technology. Alternatively, the separable state is also at the root of the power of quantum computing. The above notion has  opened up new avenues for the identification of  new quantum correlation measures beyond entanglement. Over the last couple of decades, a great deal of attention has been devoted to capture the quantumness beyond entanglement using different measures such as  quantum discord \cite{Ollivier}, measurement-induced nonlocality \cite{Luo2011MIN}, measurement-induced disturbance \cite{Luo2008MID} and skew information measures \cite{Girolami2013,UIN}. In general, the entropic measures are quite hard to compute even in  simpler systems and is shown that the computation of discord is an nondeterministic polynomial problem \cite{Huang}.

To overcome the computational complexity, the criterion of discord modified by the distance between quantum states in the state space  is known as geometric discord \cite{Dakic2010}. Different versions of geometric discord have been identified using various distance measures such as p-norm \cite{Dakic2010,trace}, Bures metric \cite{Bures} and affinity \cite{Muthuaff1}. Alternatively, Luo and Fu identified a new version of quantum correlation measure in terms of Hilbert-Schmidt (HS) distance between the state and locally perturbed state known as measurement-induced nonlocality (MIN) \cite{Luo2011MIN}. It is one such measure which captures correlation beyond entanglement and is dual to the the geometric quantum discord \cite{Dakic2010}. Due to noncontractivity of HS \cite{Piani2012},  MIN is not a faithful quantifier of bipartite quantum correlation. Till date, different versions of MIN have been introduced to resolve the above local ancilla problem \cite{Hu2012,Li2016,FMIN}. From a practical perspective, MIN is considered as a more secure resource for information processing like quantum communication and cryptography \cite{Application1,Application2,Application3,Application4}. In addition,  intrinsic decoherence \cite{Bhuvana1,Naveena1} and local noises \cite{Bhuvana2} do not seem to impact MIN.  This property can also be employed to identify quantum phase transitions \cite{Bhuvana3}. 
The realization  and implementation of the quantum algorithms constitute an  important task in the current and near-future quantum hardware which is considered to be  susceptible to  thermal fluctuations. The above challenge provides a key understanding for the development of quantum technology. Quantum magnetic materials have  the potential to be beneficial in a variety of applications like storage, magnetic sensors, medical appliances and quantum applications. The Heisenberg spin model is one of the simplest systems that possesses the nonlocal aspects. In fact, in the last two decades, the entanglement and quantum information processing in different spin models have been studied. Among them, the study of quantum correlations in mixed spin  Heisenberg models caught wide attention within the  framework of quantum information theory.  The analysis of quantum correlation in mixed spin-$(1/2, S>1)$ has been  carried out in different perspectives bringing out the impact  of uniaxial-single ion anisotropy, magnetic field and  Dzyaloshinskii–Moriya interaction \cite{HVargova2021,Sun2009,Hagiwara1999,Vargova2022,Sun2006, Hao2007, Guo2007, Wang2008, Huang2008, Li2012, Xu2014, Zhou2015, Zhou2016, Guo2014}. In addition, the role of different magnitudes of spin has  also been investigated. \cite{Hao2007,Huang2008,Zhou2015,Zhou2016,Guo2014}. The effect of g-factor on entanglement which arises due to  the noncommutativity of the magnetic moment operator and Hamiltonian of the system \cite{gfactor} has also been brought out.

Motivated by the identification of  heterodinuclear complex $[\text{Ni(dpt)}(H_2O)\text{Cu(pba)}]\cdot 2H_2 O$ [pba =1,3-propylenebis(oxamato) and dpt = bis-(3-aminopropyl)amine] which gave rise to the     experimental realization of the mixed spin-$(1/2, 1)$ Heisenberg dimer, we investigate its quantum correlations through the theoretical model of the mixed spin-$(1/2,~1)$ Heisenberg spin chain.   Its  magnetic properties have been studied  extensively \cite{Hagiwara1999, result2}.  The impact of single-ion anisotropy on entaglement has also been brought out in the interacting  mixed spin Heisenberg  model \cite{Vargova2022}.  We employ HS-MIN and fidelity-based MIN as quantifiers of quantum correlation and compare our results with that of  entanglement obtained in   \cite{PRBMixed}. Under  suitable parametric condition, it is shown that the quantum correlation measure can increase with the magnetic field in an unconventional way.  The quantum correlation between the spins decreases monotonically with the increase of temperature and uni-axial anisotropy strengthens  quantum correlations. Choosing  appropriate experimental parameters,  we also study the quantum correlation between Cu and Ni magnetic ions. Interestingly, F-MIN survives even at room temperature  $T=300K$. Further, the role of gyromagnetic $g$ factors is also highlighted. 

The paper  is organized as follows. First, we provide an overview of different measures of quantum correlations in Sec. \ref{Sec2}. Then, we introduce the  model under investigation and its diagonalization in Sec. \ref{Sec3}. We then discuss the quantum correlations in mixed spin system in Sec. \ref{Sec41}. In  Sec. \ref{Sec42}, we study the quantum correlations in CuNi complex.  Finally, in Sec. \ref{concl}, we conclude the main results of this paper.
 
\section{Measurement-induced nonlocality}\label{Sec2}
In this paper, we employ two different versions of measurement-induced nonlocality  such as Hilbert-Schmidt MIN and fidelity-based MIN pertaining to bipartite quantum correlation.  In the last decade,  researchers have identified different versions of quantum correlation measures that are based on metrics and due to the symmetry property of the Hilbert space, these measures are having merit in the computation. In this series, a new version of quantum correlation measure has been identified from the perspective of eigenprojective measurements and it is defined as \cite{Luo2011MIN}
\begin{equation}
 N_2(\rho ) =~^{\text{max}}_{\Pi ^{a}}\| \rho - \Pi ^{a}(\rho )\|_2 ^{2},
\end{equation}
where $\|\mathcal{O}\|_2^{2}=\text{Tr}(\mathcal{O}\mathcal{O}^{\dagger})$ is Hilbert-Schmidt norm of operator $\mathcal{O}$ and the maximum is taken over the locally invariant projective measurements on subsystem $a$ which does not change the state $\rho^a$. The post-measurement state is defined as $\Pi^{a}(\rho) = \sum _{k} (\Pi ^{a}_{k} \otimes   \mathds{1} ^{b}) \rho (\Pi ^{a}_{k} \otimes    \mathds{1}^{b} )$ and  $\Pi ^{a}= \{\Pi ^{a}_{k}\}= \{|k\rangle \langle k|\}$ being the projective measurements on the subsystem $a$, which do not change the marginal state $\rho^{a}$ locally i.e., $\Pi ^{a}(\rho^{a})=\rho ^{a}$. If $\rho^{a}$ is a non-degenerate state, then, the maximization is not required and the above quantity is equal to geometric discord.   An arbitrary bipartite density matrix $\rho$ can be written as 
\begin{align}
\rho=\sum_{ij}\gamma_{ij} X_i \otimes Y_j 
\end{align}
where $\gamma_{ij}=\text{Tr}(\rho(X_i\otimes Y_j))$. In a bipartite state space, the orthonormal operators in respective state spaces are $\{X_0,X_1,X_2,X_3 \}=\{\mathds{1},\sigma_1,\sigma_2,\sigma_3 \}/\sqrt{2} $ and $\{Y_0,Y_1,Y_2,Y_3 \}=\{\mathds{1},\sigma_1,\sigma_2,\sigma_3 \}/\sqrt{2} $, where $\sigma_i$ are the Pauli matrices. The above state can be recast as 
\begin{equation}
\rho=\frac{1}{4}\left[ \mathds{1}^a \otimes \mathds{1}^b+\sum_{i=1}^3 x_i (\sigma_i\otimes \mathds{1}^b)+\sum_{j=1}^3 y_j (\mathds{1}^a \otimes \sigma_j)+\sum_{i,j=1}^3 t_{ij} \sigma_i \otimes \sigma_j  \right]
\label{Equations} 
\end{equation}
where $x_i=\text{Tr}(\rho(\sigma_i\otimes \mathds{1}^b)$ and  $y_j=\text{Tr}(\rho(\mathds{1}^a \otimes \sigma_j))$ are the components of Bloch vector with  $t_{ij}=\text{Tr}(\rho( \sigma_i \otimes \sigma_j ))$ being real matrix elements of correlation matrix $T$.  MIN has a closed formula as 
\begin{equation}
N_2(\rho ) =
\begin{cases}
\text{Tr}(TT^t)-\frac{1}{\| \textbf{x}\| ^2}\textbf{x}^t TT^t\textbf{x}& 
 \text{if} \quad \textbf{x}\neq 0,\\
 \text{Tr}(TT^t)- \lambda_{\text{min}}&  \text{if} \quad \textbf{x}=0
\end{cases}
\label{HSMIN}
\end{equation}
where $\lambda_{\text{min}}$ is the least eigenvalue of matrix $TT^t$, the superscript $t$ stands for the transpose and the vector $\textbf{x}=(x_1,x_2,x_3)^t$.
 
As mentioned earlier, Hilbert-Schmidt distance is not a bonafide measure as indicated by Piani. This issue can be circumvented by modifying the definition of MIN using some other distance measures. One such quantity is fidelity-based MIN (F-MIN) and is defined as \cite{FMIN}
\begin{align}
  N_{\mathcal{F}}(\rho)=1-~^{\text{min}}_{\Pi^a}\mathcal{F}(\rho, \Pi^a(\rho)).
\end{align}
where $\mathcal{F}(\rho,\sigma)$ is the fidelity between the states \cite{Fidelity}
\begin{align}
  \mathcal{F}(\rho,\sigma)=\frac{\text{Tr}(\rho\sigma)^2}{\text{Tr}(\rho^2)\text{Tr}(\sigma^2)},
\end{align} 
which can be computed easily  compared to fidelity introduced by Josza \cite{Jozsa1994} and satisfies all the properties of a good measure of fidelity between states. In addition, one can realize the above fidelity using quantum circuits \cite{subf}. Here also, the minimization is taken over the locally invariant projective measurements. Due to the multiplicative property of the fidelity, F-MIN fixes the local ancilla problem. The closed formula of F-MIN is computed as 
\begin{equation}
N_{\mathcal{F}}(\rho ) =
\begin{cases}
\text{Tr}(\Gamma\Gamma^t)-TrA\Gamma\Gamma^tA^t & 
 \text{if} \quad \textbf{x}\neq 0,\\
 \text{Tr}(\Gamma\Gamma^t)- \tau_{\text{min}}&  \text{if} \quad \textbf{x}=0
\end{cases}
\label{F-MIN}
\end{equation}
where $\tau_{\text{min}}$ is the minimal eigenvalue of the matrix $\Gamma\Gamma^t$ and the matrix A is given by
\begin{align}
A=\frac{1}{\sqrt{2}}
\begin{pmatrix}
  1 & \frac{\textbf{x}}{|\textbf{x} |} \\
  1 & -\frac{\textbf{x}}{|\textbf{x} |}
\end{pmatrix}.
\end{align}
\section{The model and thermalization}\label{Sec3}
To understand the behaviors of thermal quantum correlations, we consider the Hamiltonian of the mixed spin - $(1/2, 1)$ Heisenberg dimer and is defined as,
\begin{align}
\mathcal{H}=J\left[ \Delta (\hat{S}^x \hat{\Sigma}^x+\hat{S}^y \hat{\Sigma}^y )+\hat{S}^z \hat{\Sigma}^z\right] + D(\hat{\Sigma}^z)^2-g_1\mu_BB \hat{S}^z-g_2\mu_B B\hat{\Sigma}^z
\end{align}
where $\hat{S}^{\alpha}(\hat{\Sigma}^{\alpha})$  denotes the spatial components of the spin-$1/2 (1)$ operators with $\alpha=x,y,z$, $J$ is the coupling constant  between the spin-1/2 and spin-1 magnetic ions, the parameter $\Delta$ determines the XXZ exchange anisotropy in this exchange interaction, $D$ is a uniaxial single-ion anisotropy acting on the spin-1 magnetic ions only, $B$ denotes a static external magnetic field, $\mu_B$ is the Bohr magneton and, $g_1$ and $g_2$ are Landé $g$ factors of the spin-1/2 and spin-1 magnetic ions respectively. 

It is worth mentioning at this juncture that the above  mixed spin-$(1/2,1)$ theoretical model is realizable in hetero-bimetallic complexes such as the CuNi compound. Here, the spin-1/2 $Cu^{2+}$ and spin-1 $Ni^{2+}$ are coupled via Heisenberg exchange coupling $J$. In the standard qubit-qutrit computational basis $\{ |\frac{1}{2},0\rangle,|\frac{-1}{2},0\rangle\,|\frac{1}{2},1\rangle,|\frac{1}{2},-1\rangle, |\frac{-1}{2},1\rangle, |\frac{-1}{2},-1\rangle\} $, the Hamiltonian has the following matrix form, 
\begin{align}
\mathcal{H} = \begin{pmatrix}
 A_- & 0 & 0 & 0 & 0 & 0 \\
 0 & B_- & 0 & \nu & 0 & 0 \\
 0 & 0 & C_+ & 0 & \nu & 0 \\
0 & \nu & 0 & C_- & 0 & 0 \\
0 & 0 & \nu & 0 & B_+ & 0 \\
0 & 0 & 0 & 0 & 0 &A_+ \\
\end{pmatrix},
\label{ham}
\end{align}

where the diagonal elements are
\begin{align}
A_{\pm}=\frac{1}{2}[J+2D\pm(h_{1}+2h_{2})],~~~~~ B_{\pm}=\pm \frac{h_{1}}{2}~~~~~ \text{and}~~~~ C_{\pm}=-\frac{1}{2}[J-2D\pm(h_{1}-2h_{2})]. \nonumber
\end{align} 
The only off-diagonal element is $\nu=J\Delta/\sqrt{2}$. 
The eigenvalues and the corresponding eigenvectors  of the Hamiltonian $\mathcal{H}$ are computed as 
\begin{eqnarray}
E_{1,2}=\frac{1}{2}[J+2D \mp (h_{1}+2h_{2})], & ~~~~~~~~~~~~~~~\vert\varphi_{1}\rangle=|\frac{1}{2},1\rangle,& \vert\varphi_{2}\rangle=|\frac{-1}{2},-1\rangle\nonumber \\
E_{3,4}=\frac{-1}{4}[J-2D+2h_{2}] \mp \frac{1}{4} \eta_{-}, & ~~~~~~~~~~~~ \vert\varphi_{3,4}\rangle=c_{1}^\mp|\frac{1}{2},0\rangle \mp c_{1}^\pm|\frac{-1}{2},1\rangle  \nonumber\\ 
E_{5,6}=\frac{-1}{4}[J-2D-2h_{2}] \mp \frac{1}{4} \eta_{+}, & ~~~~~~~~~~~~~~~ \vert\varphi_{5,6}\rangle=c_{2}^\pm|\frac{1}{2},-1\rangle \mp c_{2}^\mp|\frac{-1}{2},0\rangle. \nonumber 
\end{eqnarray}
The normalization constants are 
\begin{align}
c_{1}^\pm=\frac{1}{\sqrt{2}}\sqrt{1 \pm \frac{J-2D-2(h_{1}-h_{2})}{\eta_{-}}} ~~~~~~~~~~\text{and}~~~~~
c_{2}^\pm=\frac{1}{\sqrt{2}}\sqrt{1 \pm \frac{J-2D+2(h_{1}-h_{2})}{\eta_{+}}}
\end{align}
with the parameter $\eta_{\pm} = \sqrt{[J-2D \pm 2(h_{1}-h_{2})]^2+8(J\Delta)^2}$. 

The thermal density matrix for the mixed spin-(1/2, 1) Heisenberg dimer is  
\begin{align}
  \varrho(T)=\frac{1}{\mathcal{Z}}\exp{\left(-\beta \mathcal{H}\right)}=\frac{1}{\mathcal{Z}}\sum_{i=1}^6 p_i \vert \varphi_{i}\rangle \langle \varphi_{i}\vert,
\end{align}

 where $\beta=1/k_BT$ and it can be calculated as
\begin{align}
\varrho(T) = \frac{1}{\mathcal{Z}}\begin{pmatrix}
 \varrho_{11} & 0 & 0 & 0 & 0 & 0 \\
 0 & \varrho_{22} & 0 & \varrho_{24} & 0 & 0 \\
 0 & 0 & \varrho_{33} & 0 & \varrho_{35} & 0 \\
0 & \varrho_{42} & 0 & \varrho_{44} & 0 & 0 \\
0 & 0 & \varrho_{53} & 0 & \varrho_{55} & 0 \\
0 & 0 & 0 & 0 & 0 & \varrho_{66} \\
\end{pmatrix},
\label{thermal}
\end{align}
where the matrix elements are
\begin{eqnarray}
\varrho_{11}={\frac{1}{Z}}\mathrm{e}^{\frac{-\beta}{2}(J+2D-(h_{1}+2h_{2}))}, ~~~~~~\varrho_{66}={\frac{1}{Z}}\mathrm{e}^{\frac{-\beta}{2}(J+2D+(h_{1}+2h_{2}))} ~~~~~~~~~~~~~~~~~~~~~~~~~~~~~~~~~~~~\nonumber \\  
\varrho_{22}={\frac{1}{Z}}\mathrm{e}^{\frac{\beta}{4}(J-2D+2h_{2})}
\left[\text{cosh}\left(\frac{\beta\eta_{-}}{4}\right)-\frac{(J-2D-2(h_{1}-h_{2}))}{\eta_{-}}\text{sinh}\left(\frac{\beta\eta_{-}}{4}\right)\right], ~~~~~~~~~~~~~~~~\nonumber \\
\varrho_{33}={\frac{1}{Z}}\mathrm{e}^{\frac{\beta}{4}(J-2D-2h_{2})}
\left[\text{cosh}(\frac{\beta\eta_{+}}{4})+\frac{(J-2D+2(h_{1}-h_{2}))}{\eta_{+}}\text{sinh}(\frac{\beta\eta_{+}}{4})\right], ~~~~~~~~~~~~~~~~~~~~\nonumber \\
\varrho_{44}={\frac{1}{Z}}\mathrm{e}^{\frac{\beta}{4}(J-2D+2h_{2})}\left[
\text{cosh}(\frac{\beta\eta_{-}}{4})+\frac{(J-2D-2(h_{1}-h_{2}))}{\eta_{-}}\text{sinh}(\frac{\beta\eta_{-}}{4})\right],~~~~~~~~~~~~~~~~~~~~ \nonumber \\
\varrho_{55}={\frac{1}{Z}}\mathrm{e}^{\frac{\beta}{4}(J-2D-2h_{2})}\left[
\text{cosh}(\frac{\beta\eta_{+}}{4})-\frac{(J-2D+2(h_{1}-h_{2}))}{\eta_{+}}\text{sinh}(\frac{\beta\eta_{+}}{4})\right],~~~~~~~~~~~~~~~~~~~~ \nonumber \\
\varrho_{24}=\varrho_{42}=\frac{-\sqrt{8}(J\Delta)}{Z\eta_{-}}\mathrm{e}^{\frac{\beta}{4}(J-2D+2h_{2})}\text{sinh}\left(\frac{\beta\eta_{-}}{4}\right),~~~~~~~~~~~~~~~~~~~~~~~~~~~~~~~~~~~~~~~~~~~~~~~~~~  \nonumber \\ 
\text{and}~~~~~~~~~~~~~~~~ ~~~~~~~~~~~~~~~~~~~~~~~~~~~~~~~~~~~~~~~~~~~~~~~~~~~~~~~~~~~~~~~~~~~~~~~~~~~~~~~~~~~~~~~~~~~\nonumber \\
\varrho_{35}=\varrho_{53}=\frac{-\sqrt{8}J\Delta}{Z\eta_{+}}\mathrm{e}^{\frac{\beta}{4}(J-2D-2h_{2})}\text{sinh}\left(\frac{\beta\eta_{+}}{4}\right).~~~~~~~~~~~~~~~~~~~~~~~~~~~~~~~~~~~~~~~~~~~~~~~~~~~~ \nonumber
\end{eqnarray}

The partition function of the system is given by
\begin{eqnarray}
\mathcal{Z}=2\left[{\mathrm{e}^{\frac{-\beta(J+2D)}{2}}\text{cosh}\left(\frac{\beta(h_{1}+2h_{2})}{2}\right)+\mathrm{e}^{\frac{\beta(J-2D)}{4}}\left[\mathrm{e}^{\frac{\beta h_{2}}{2}}\text{cosh}\left(\frac{\beta\eta_{-}}{4}\right)+\mathrm{e}^{\frac{-\beta h_{2}}{2}}\text{cosh}\left(\frac{\beta\eta_{+}}{4}\right)\right]}\right].
\end{eqnarray}


\section{Results and Discussions}\label{Sec4}
In this section, we study the bipartite thermal quantum correlations quantified by  MIN and fidelity-based MIN (F-MIN) of the  mixed spin Heisenberg dimer and compare with that of entanglement. Using  Eqs. (4), (7), and elements of the density matrix analytically, we  have computed  MIN and F-MIN of the above the mixed spin system. It can be recalled that the entanglement of  the above physical system is already studied in Ref.\cite{PRBMixed}. 
\subsection{Quantum correlation in thermal states}
\label{Sec41}
\begin{figure*}[!ht]
\centering\includegraphics[width=0.45\linewidth]{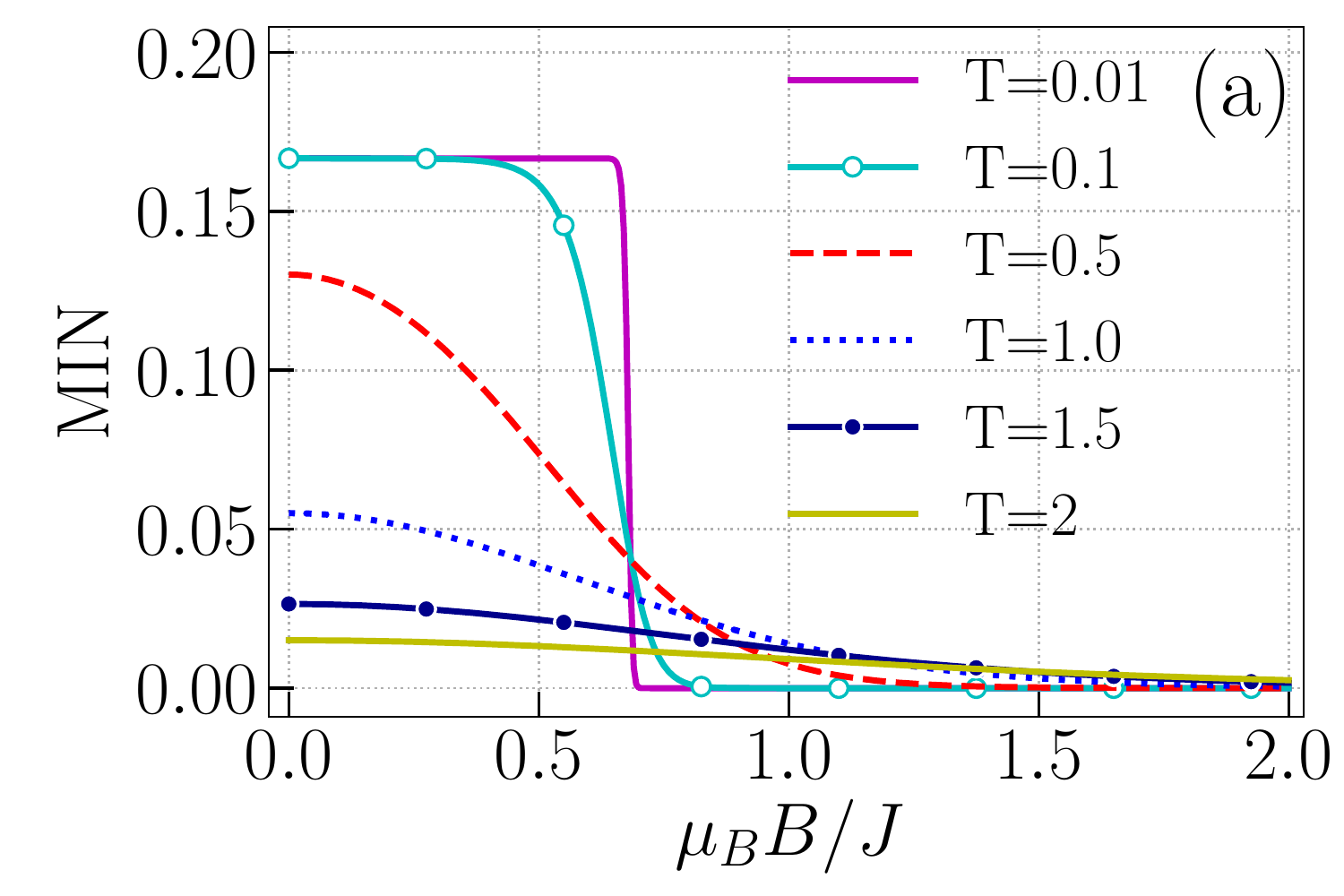}
\centering\includegraphics[width=0.45\linewidth]{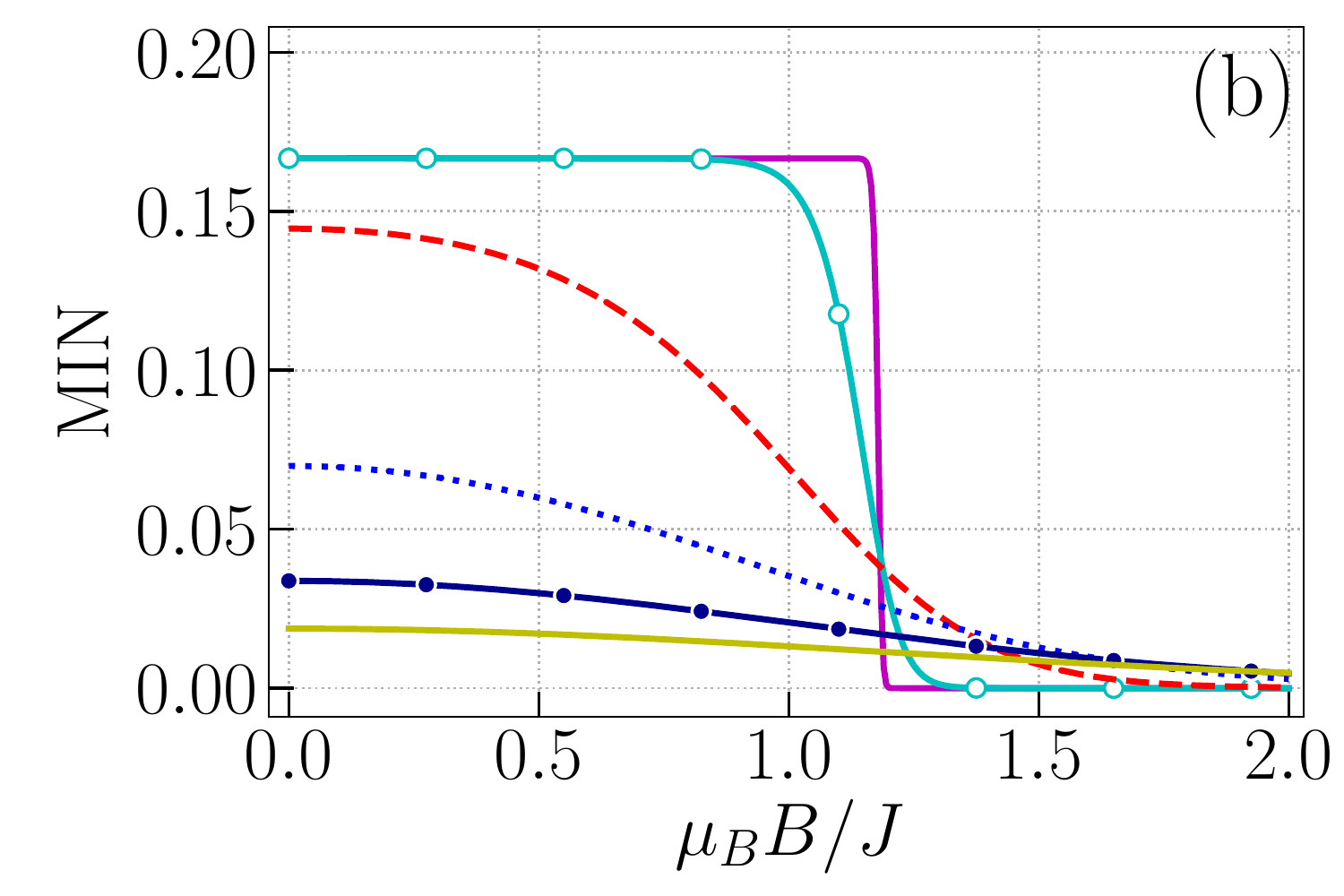}
\centering\includegraphics[width=0.45\linewidth]{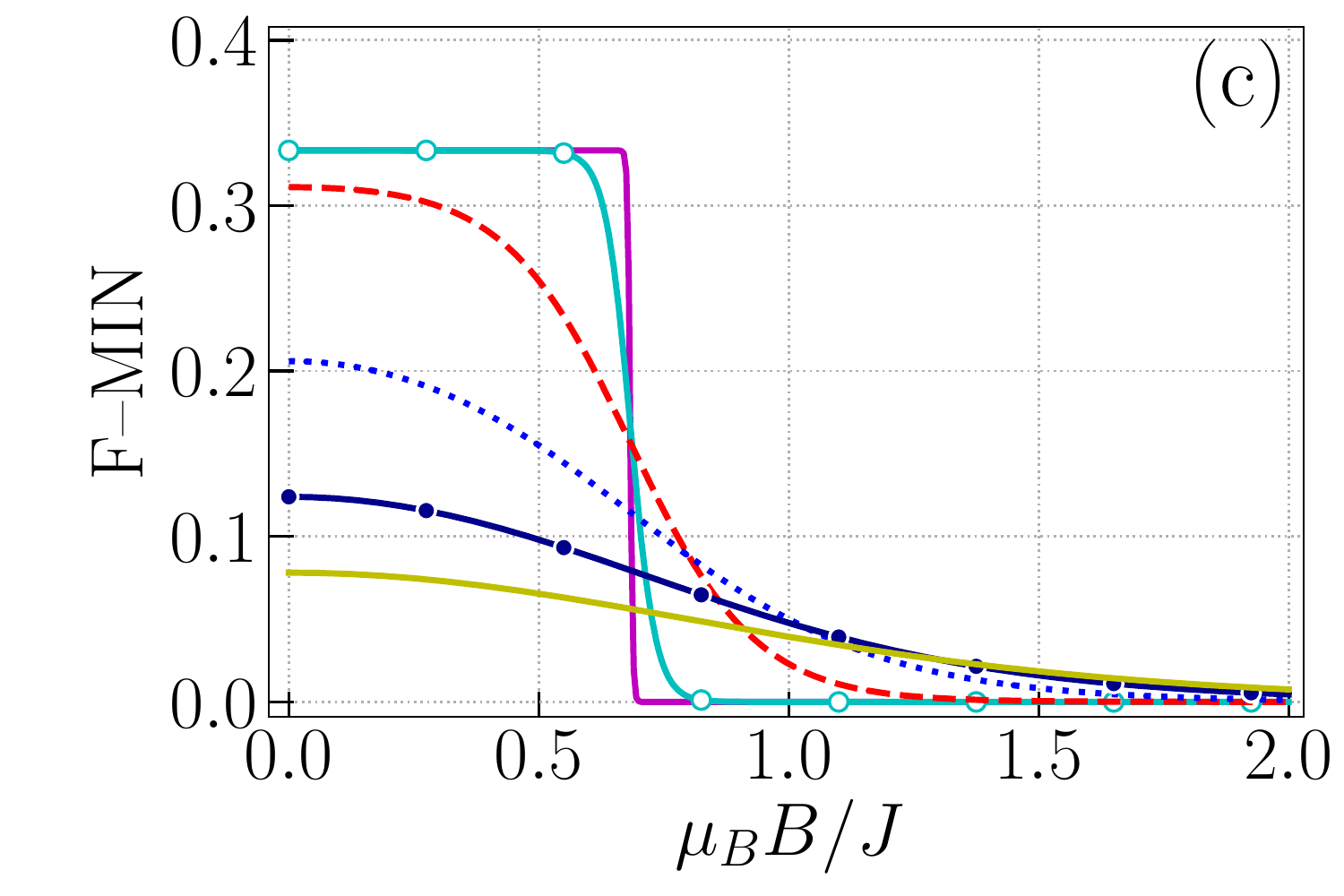}
\centering\includegraphics[width=0.45\linewidth]{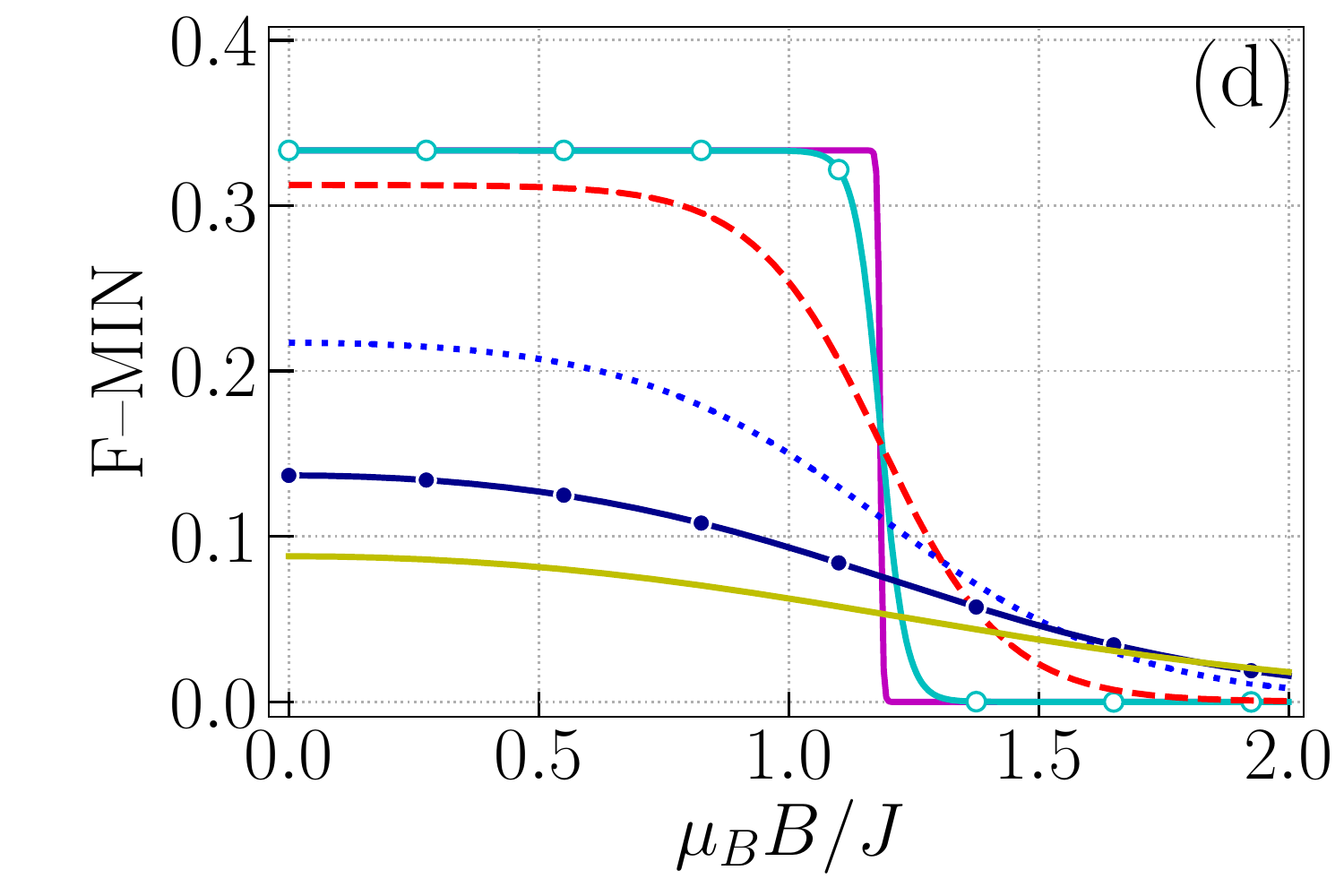}
\caption{ Thermal quantum correlations quantified by (a) \& (b) MIN  and (c) \& (d)  F-MIN as a function of external magnetic field  for  different temperatures. The fixed parameters $ g_1 = g_2 = 2, \Delta = 1, (a)~ \&~ (c)~ D/J = -0.5, (b) ~\&~ (d)~ D/J = 1.5 $.}

\label{Fig1}
\end{figure*}

First, we study the influence of the magnetic field on quantum correlation for  different  temperatures for two different values of the uniaxial single-ion anisotropy such as $D/J=-0.5 ~\text{and}~1.5$.   For this purpose, we fix the values $g_1=g_2=2$, $\Delta=1$ .  In general, the magnetic field and thermal fluctuations suppress the degree of quantumness in the interacting spin systems. In Fig. \ref{Fig1}, we illustrate the behaviors of quantum correlation measures as  a function of the magnetic field for a fixed single ion anisotropy parameter. Initially, we observe that the MINs remain constant when we  increase the magnetic field at sufficiently  low temperatures, then drop to zero at a critical magnetic field. On the other hand, the entanglement initially increases  unconventionally  with the magnetic field due to Zeeman’s splitting of two energy levels \cite{PRBMixed}. 

To understand the role of temperature on MINs, they are plotted for different temperatures in Fig. \ref{Fig1}. In this connection, we observe that increase of temperature causes monotonic decrease of quantum correlation from the maximum value  to zero in the mixed spin system.  At higher temperatures, the correlation between the spins is very small and decreases with the increase of magnetic field. In addition, the comparison of  Fig. \ref{Fig1}(a) and Fig. \ref{Fig1}(b) drives home the point  that the nonzero MIN region increases with the increase of the single-ion anisotropy parameter $D$. One observes a  similar functional behavior for F-MIN.

\begin{figure*}[!ht]
\centering\includegraphics[width=0.3\linewidth]{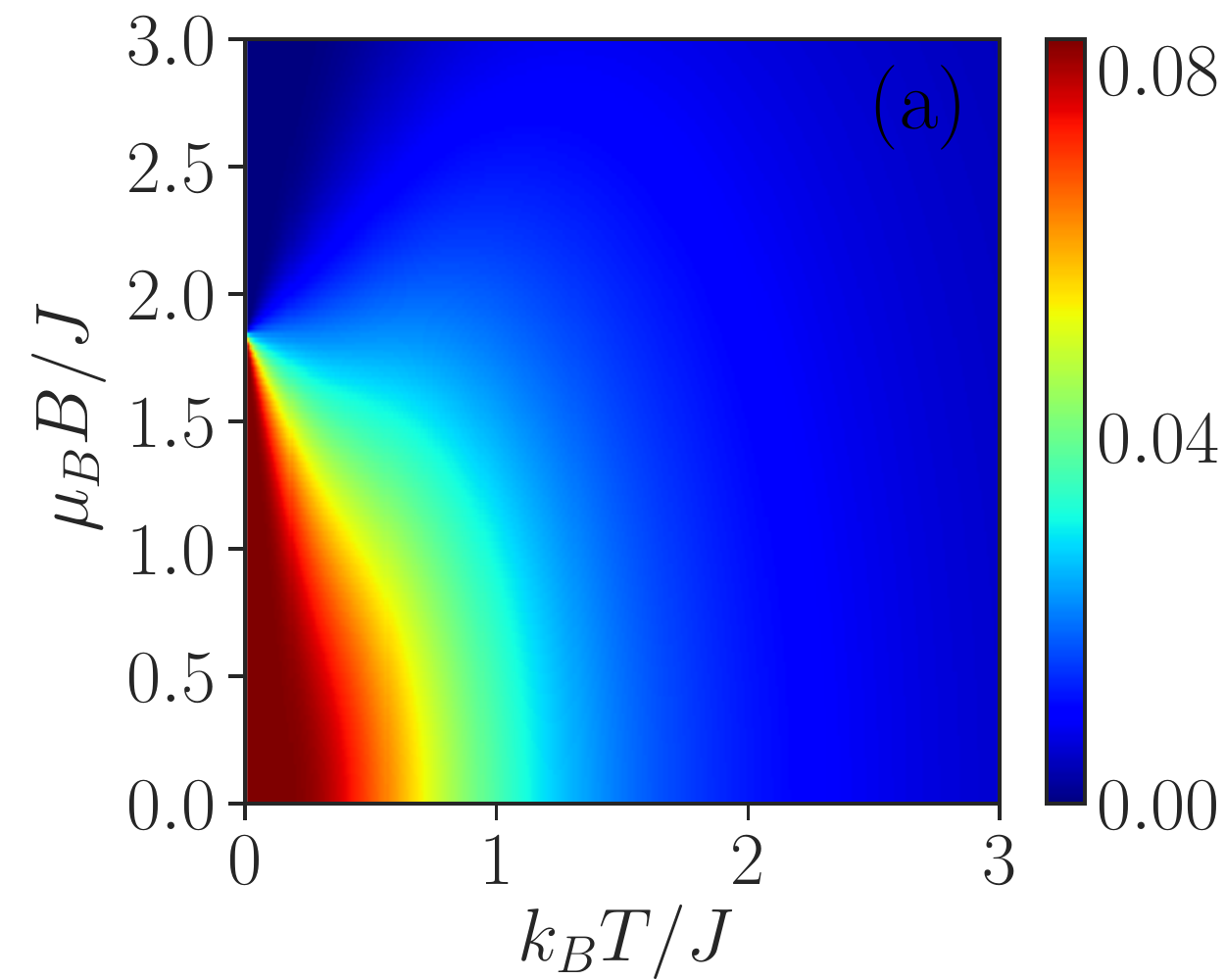}
\centering\includegraphics[width=0.3\linewidth]{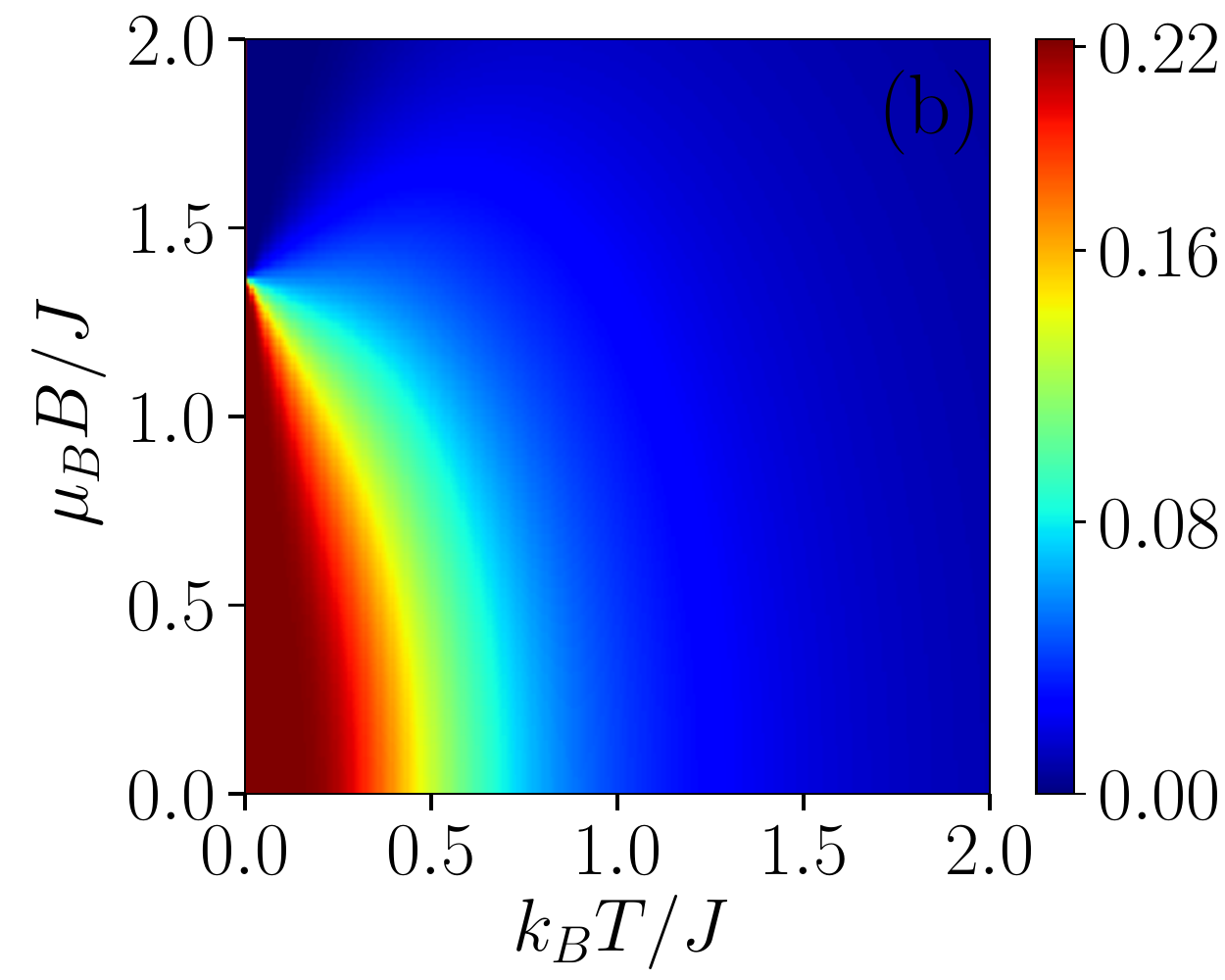}
\centering\includegraphics[width=0.3\linewidth]{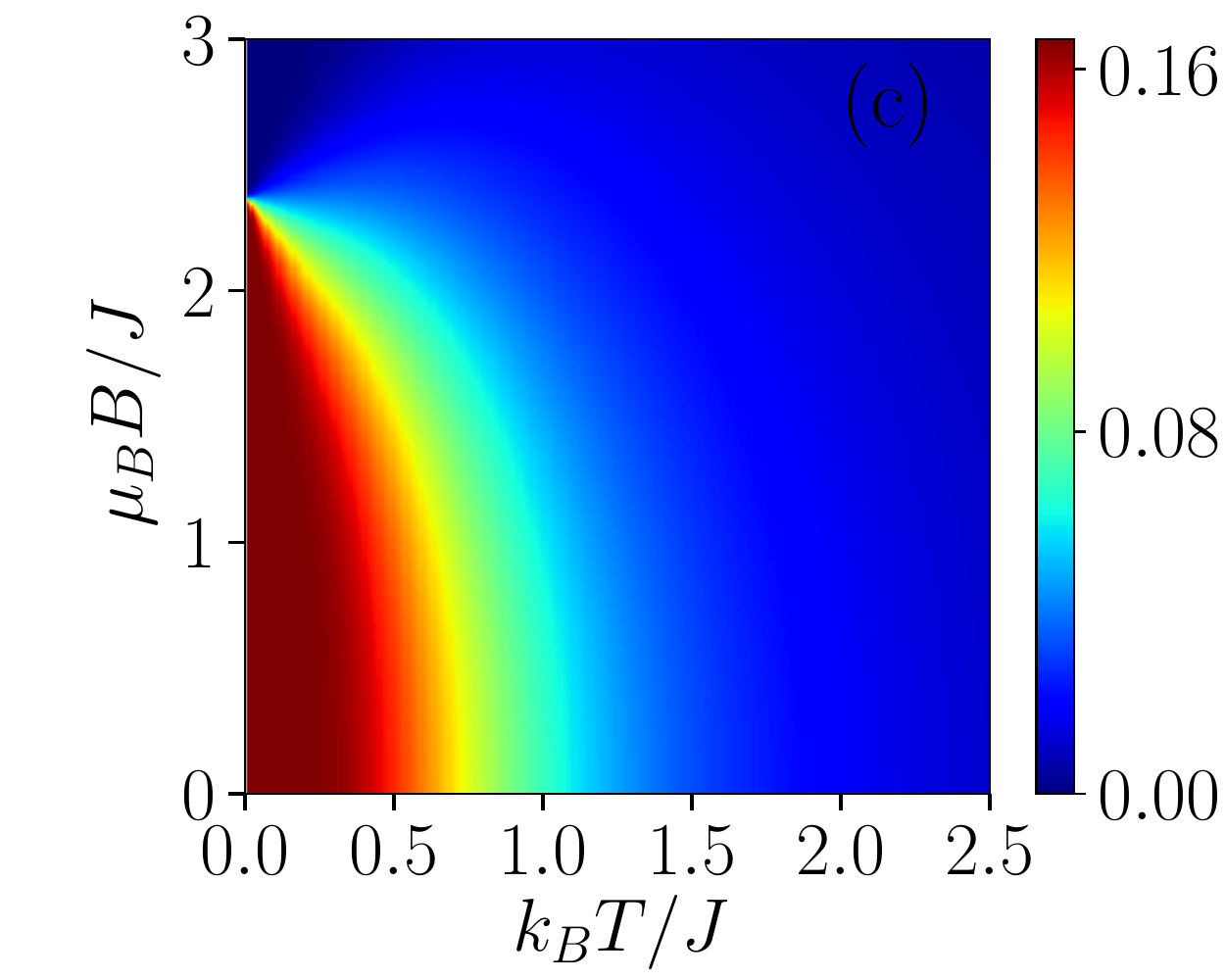}
\centering\includegraphics[width=0.3\linewidth]{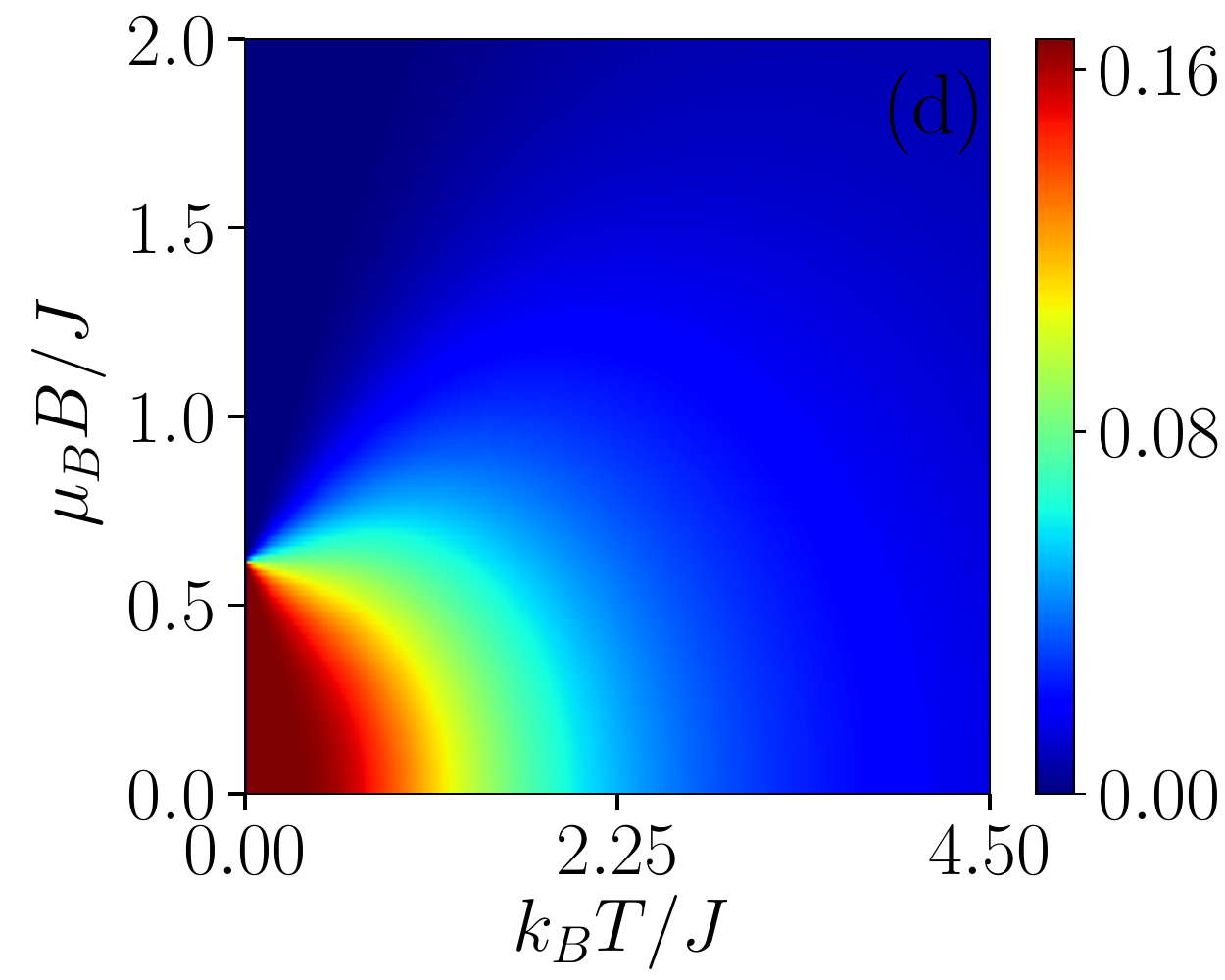}
\centering\includegraphics[width=0.3\linewidth]{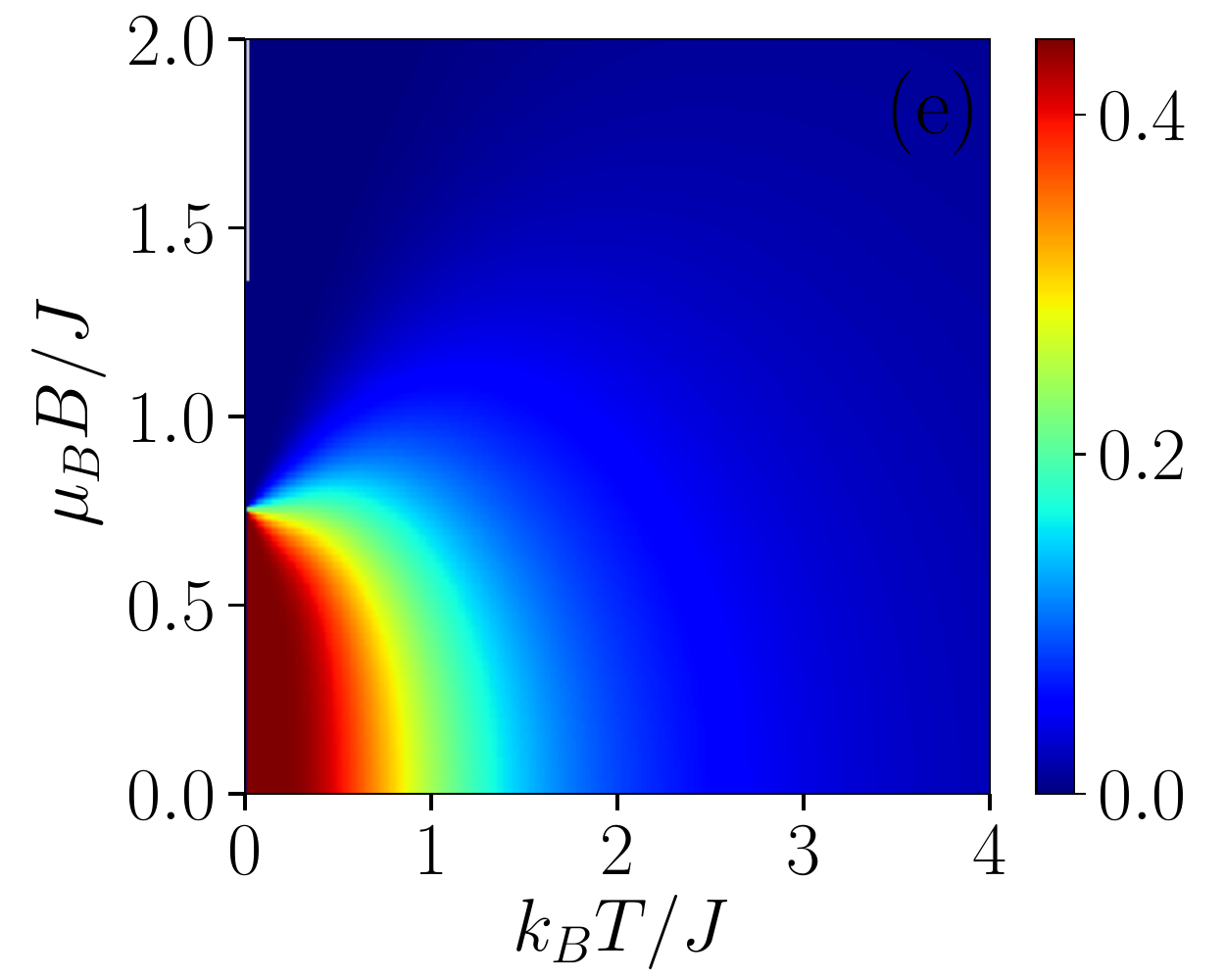}
\centering\includegraphics[width=0.3\linewidth]{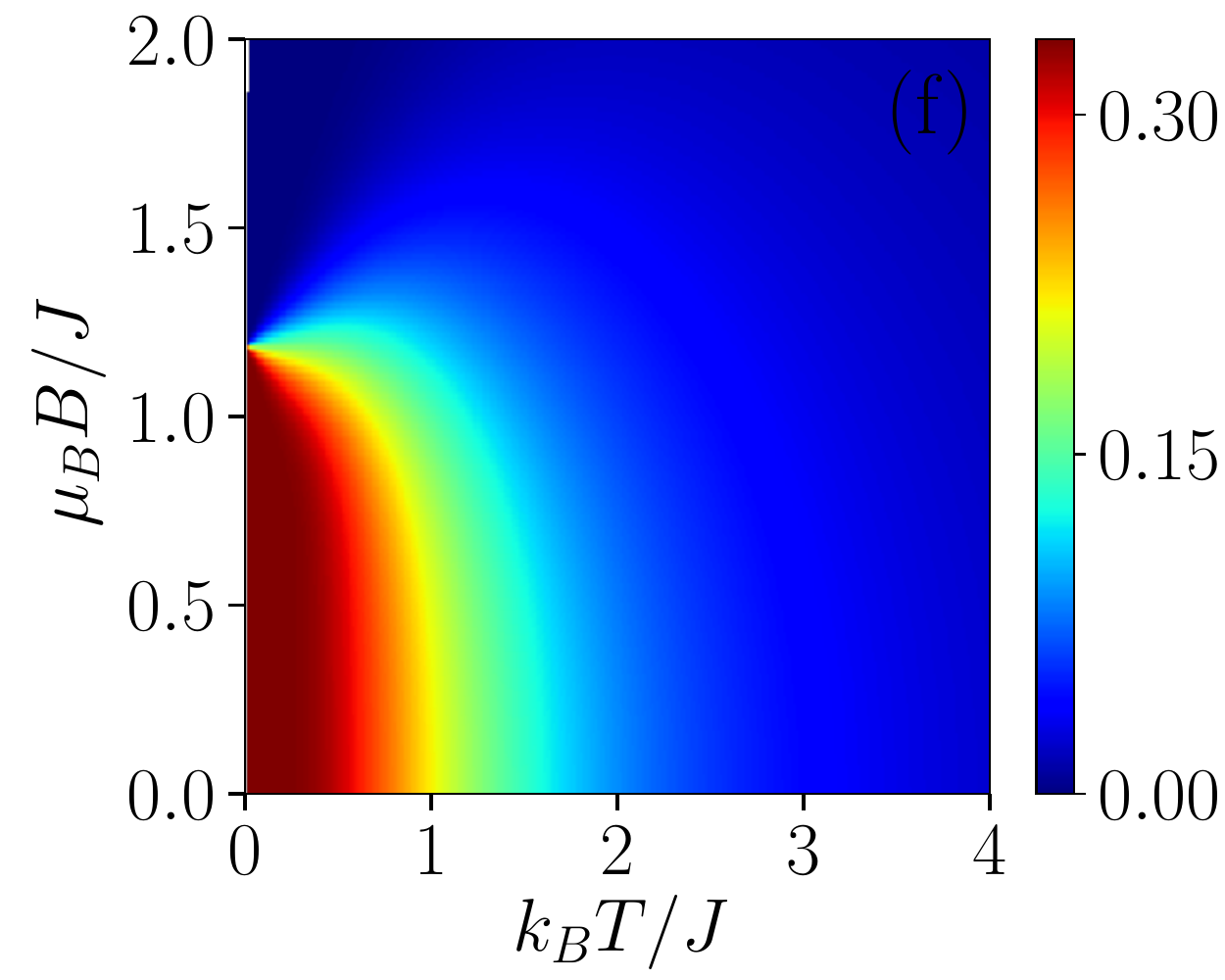}
\caption{Density of MIN (upper panel) and F-MIN (lower panel) as a function of external magnetic field and temperature (a) \& (d) $D/J =-1.5$, (b) \& (e) $D/J =0$, (c) \& (f) $D/J =1.5$. The fixed parameters are $g_1 = g_2 = 2$ and $\Delta=1$.}
\label{Fig2}
\end{figure*}

 To attain a deeper insight into the influence of uniaxial single-ion anisotropy $D$ on the thermal quantum correlation, in  Fig. \ref{Fig2}, we plot the densities of MIN as a function of magnetic field and temperature for a given value of $g$-factors. Like entanglement,  here also, we observe that the spins are strongly correlated at low temperature and weak magnetic field regions. It is pretty obvious that  MINs  capture more quantumness between the spins unlike entanglement. Further, we observe that the parameter $D$ increases the quantum correlation  implying that the enhancement of $D$ can increase the threshold values of temperature and external magnetic field. In other words, the parameter $D$ introduces the correlation between the spins. Further, we notice  that the single-ion anisotropy induces the correlation in the parametric space where there is no correlation between the spins and strengthens the correlation in the parametric space if the spins are already correlated similar to $DM$ interaction \cite{Bhuvana3}.   In the asymptotic limit, all eigenstates would be given by eigenvectors with a single separable basis state vector without any quantum superposition, i.e., without any quantum correlation. This  implies that the increase of quantum correlation by single-ion anisotropy $D$ is not a generic feature. 

\begin{figure*}[!ht]
\centering\includegraphics[width=0.45\linewidth]{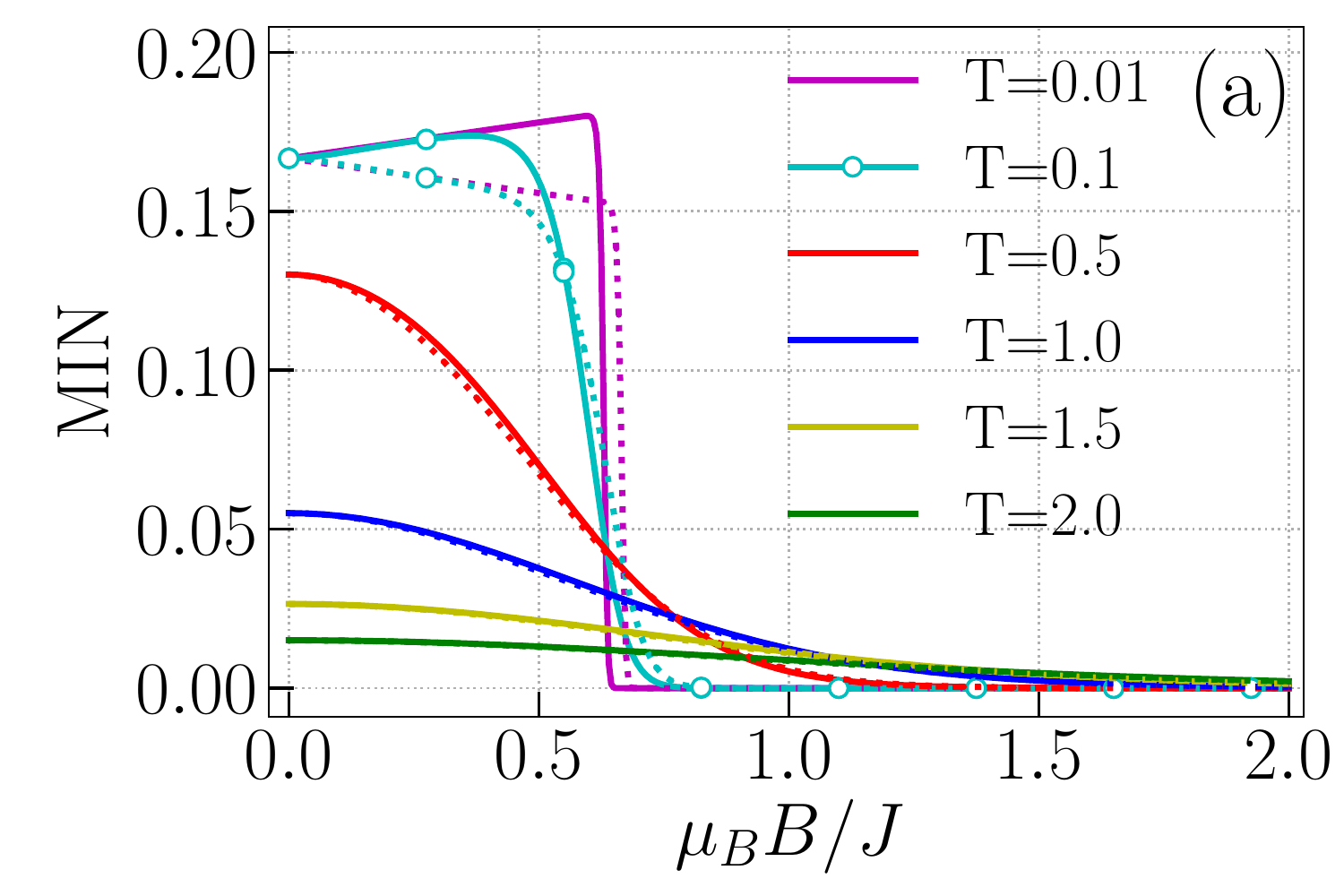}
\centering\includegraphics[width=0.45\linewidth]{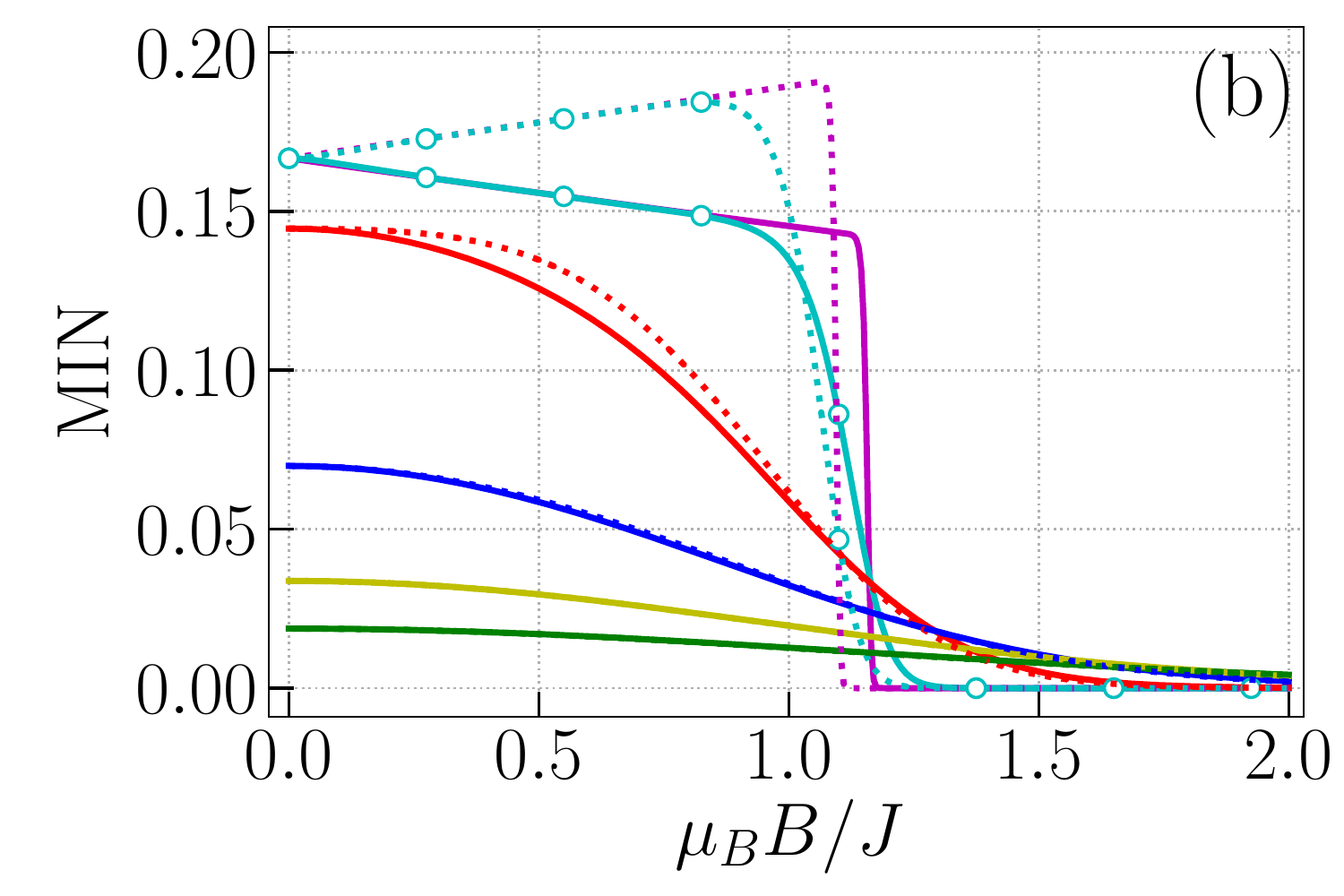}
\centering\includegraphics[width=0.45\linewidth]{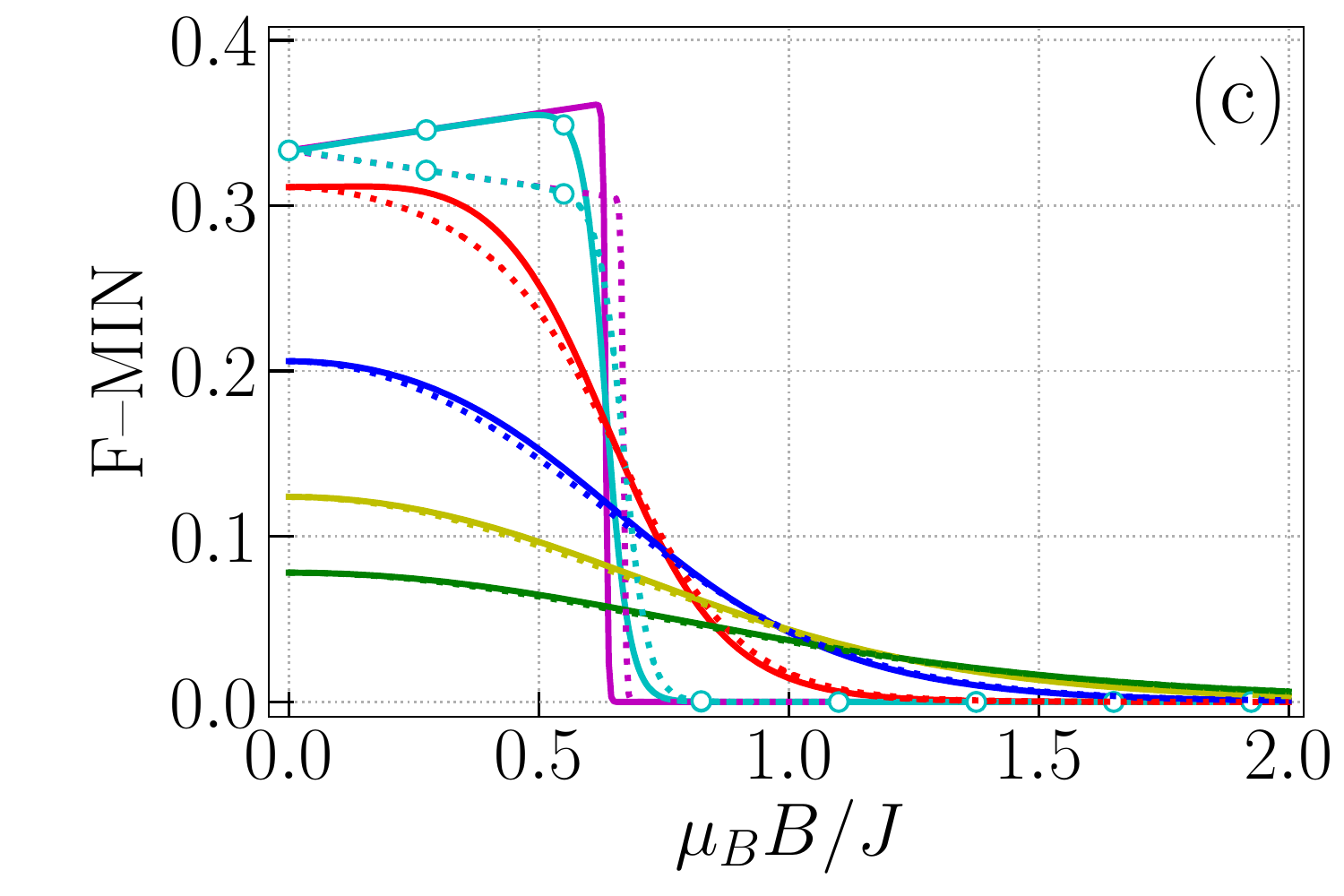}
\centering\includegraphics[width=0.45\linewidth]{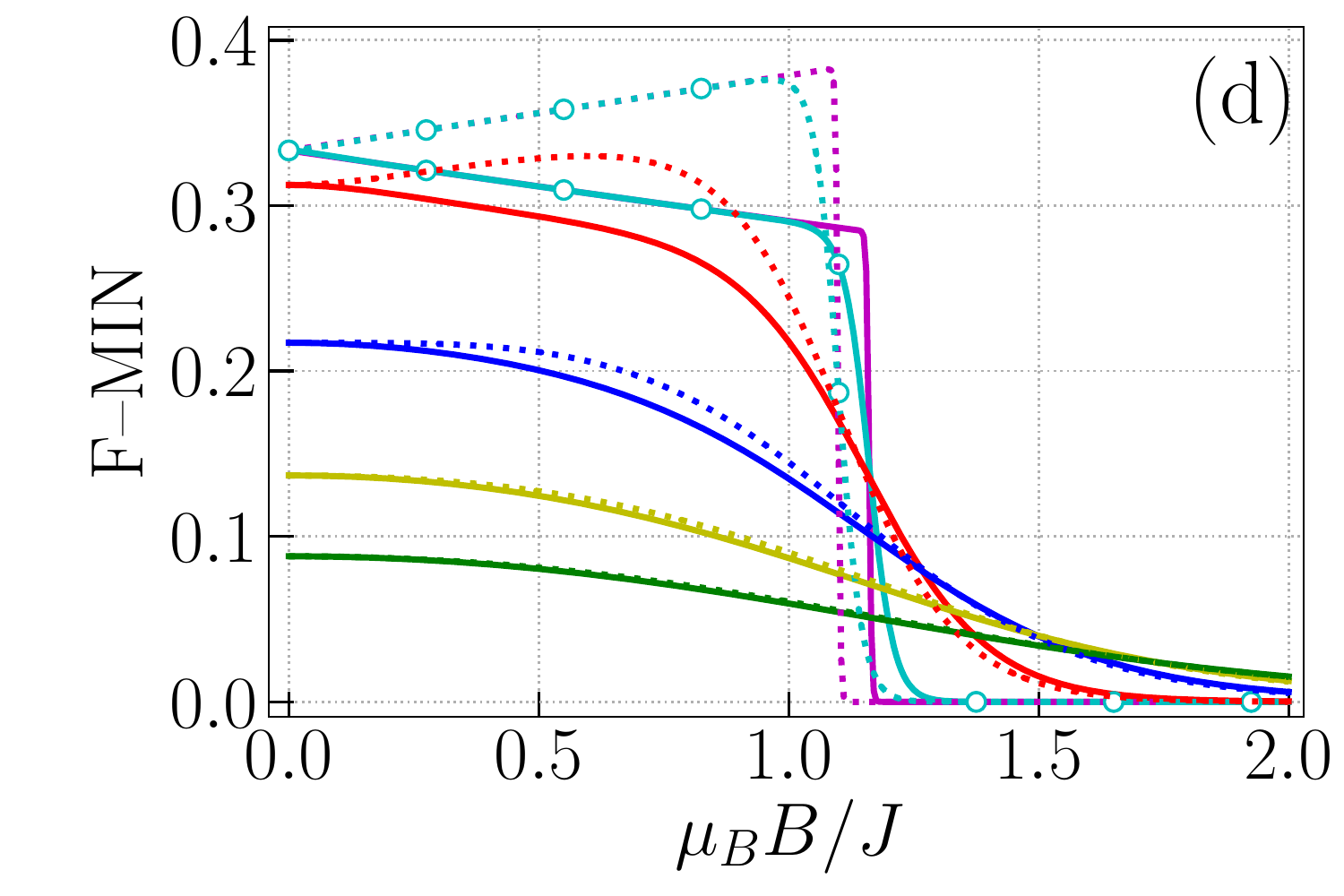}
\caption{ Thermal quantum correlations quantified by MIN and F-MIN as a function of external magnetic field (a) \& (c) $D/J=-0.5$ (b) \& (d) $D/J=1.5$ for  different temperatures. The fixed parameters  $g_1 = g_2=2.2$ and  $\Delta=1$.}


\label{Fig3}
\end{figure*}

Next, we analyze the impact of difference of the $g-$factors on quantum correlations.  Here, we consider the difference of the g-factors $|g_1-g_2 | =0.2$ under two criteria such as $g_1< g_2$ and $g_1> g_2$. To understand the effects of g-factor on MINs, we have plotted the measures as a function of magnetic field for different temperatures for a given single-ion anisotropy parameter. Even for a small difference of g-factors, the contribution  of g-factor turns out to be  subtle. For  $g_1<  g_2$ ( $g_1 > g_2$ ), the behaviors of MIN and F-MIN are illustrated in solid (dashed) lines in  Fig. \ref{Fig3}. When $g_1<  g_2$  and $D=-0.5$, we notice that the measures increase unconventionally with magnetic field at low temperature $T=0.1$. For  $g_1< g_2$, MINs are found to decrease at   $T=0.1$. The significant role of difference of g-factors is more pronounced at  low temperatures and the increase of temperatures negates  the impact of difference of g-factors.  Nevertheless, it should be  pointed out that the negativity tends to the same asymptotic value in the zero-field limit as well as at high magnetic fields while the most pronounced differences can be  detected at low magnetic fields. Again, it can be noticed that   the single-ion anisotropy parameter induces and strengthens the quantum correlation between the mixed spins system.

\subsection{ Quantum Correlation in CuNi Complex}
\label{Sec42}
In this section, we study the quantum correlation of the heterodinuclear complex CuNi. It is worth pointing at this juncture  that the magnetic properties of the above mixed spin (1/2,1) Heisenberg dimer have been  experimentally predicted through the CuNi compound and verified \cite{Hagiwara1999,result2}. In the following, we  therefore invoke the same set of  model parameters  to make the relevant theoretical prediction for the bipartite entanglement of the CuNi dimeric compound. The reported parameter values of the CuNi compound are $J/k_B=141 K$ and g-factors of $\text{Cu}^{2+}$ and $\text{Ni}^{2+}$ are $g_1=2.20$ and $g_2=2.29$ respectively  \cite{Hagiwara1999}. Similarly, the other    parametric values reported are $J/k_B=121 K$, $g_1=2.09$ and $g_2=2.22$ \cite{result2}. Here, we use the experimental results given in \cite{Hagiwara1999}.

\begin{figure*}[!ht]
\centering\includegraphics[width=0.45\linewidth]{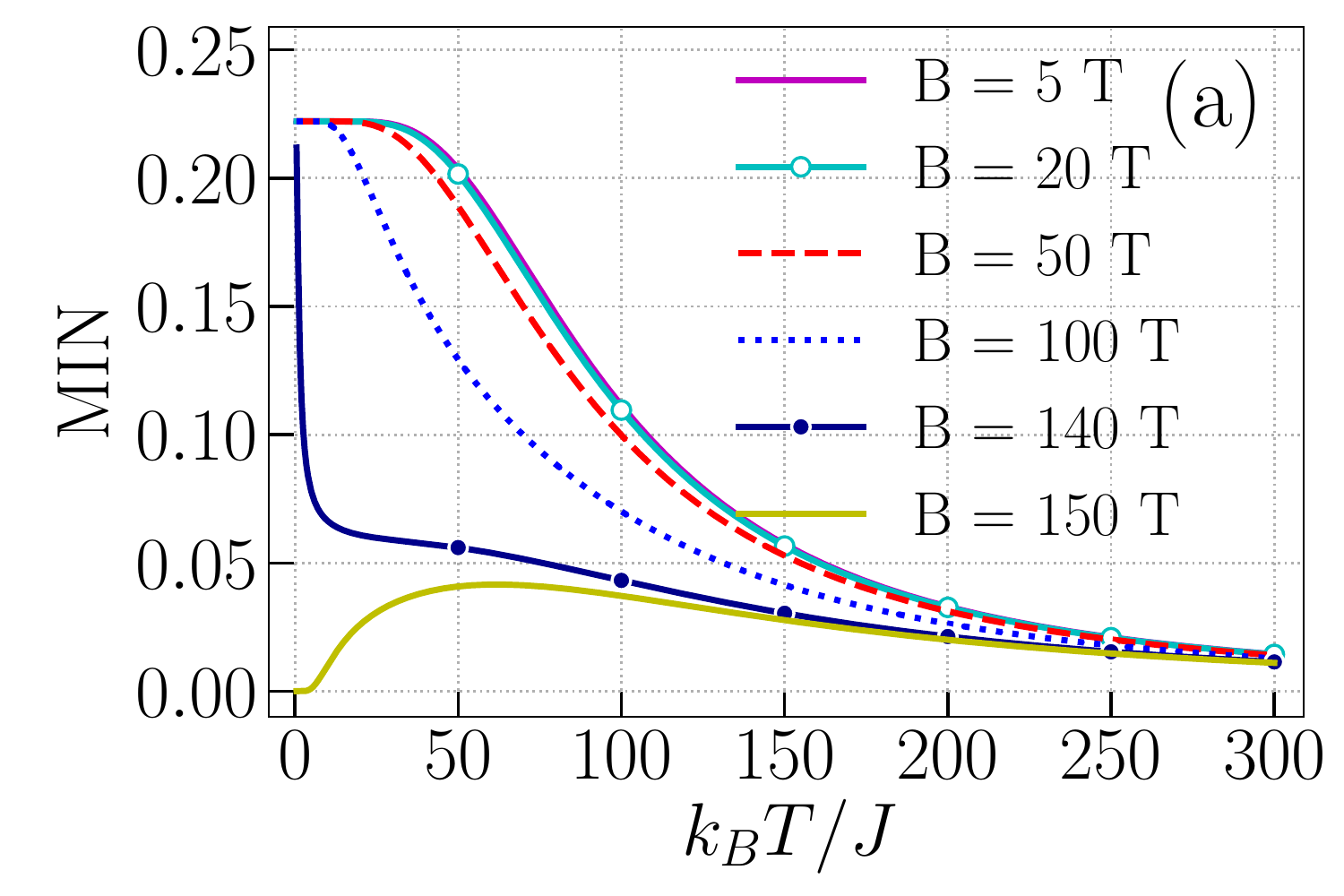}
\centering\includegraphics[width=0.45\linewidth]{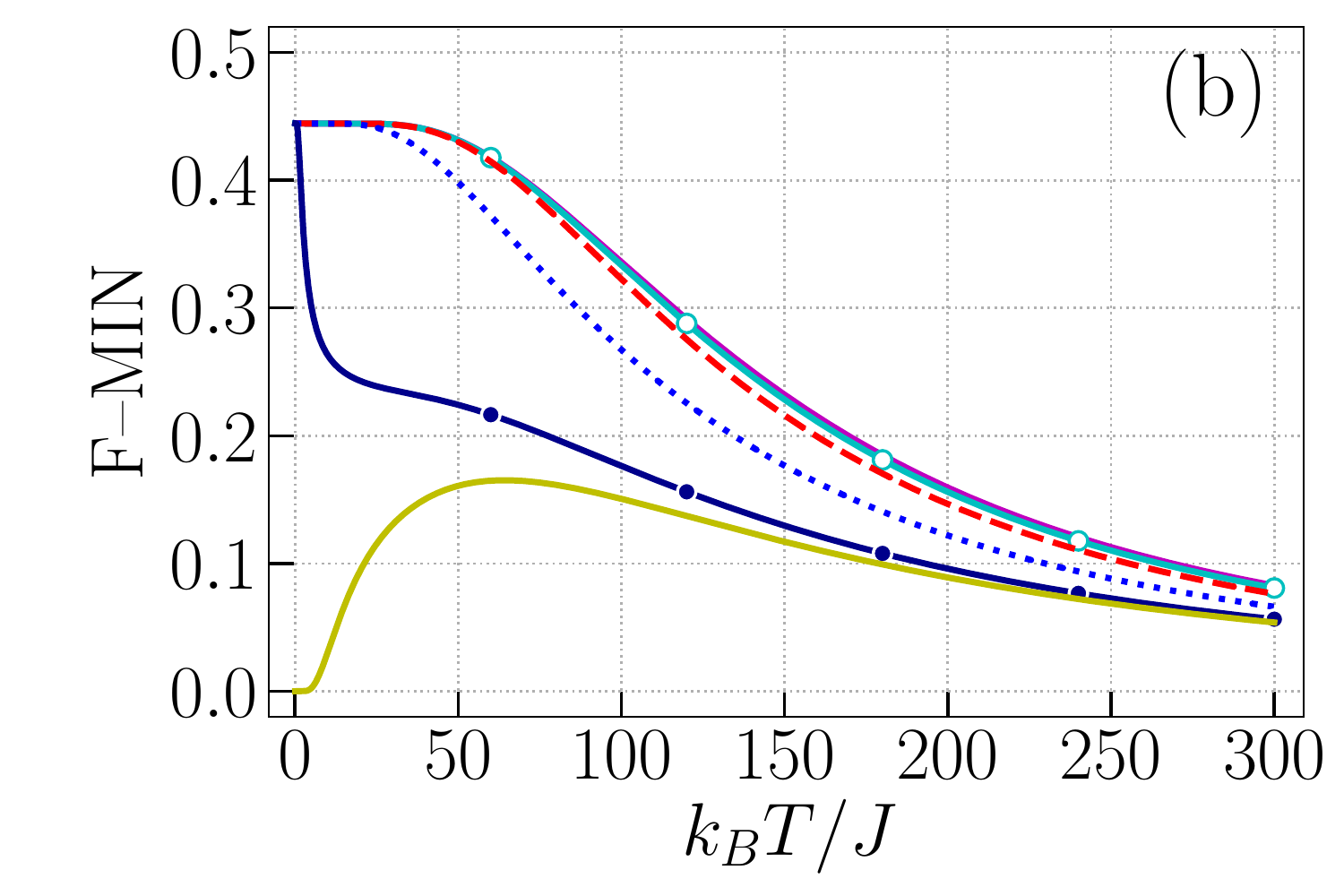}
\caption{ Thermal quantum correlations quantified by MIN and F-MIN as a function of temperature and for  different external magnetic field. The fixed parameters are  $g_1 =2.2$,  $g_2=2.29$, $J/k_B=141 K$, ~ and ~$ D/k_B = 0$.}

\label{Fig4}
\end{figure*}
We measure theoretically quantum correlation quantified by MIN and F-MIN in CuNi complex through the mixed spin (1/2,1) Heisenberg dimer. Figure. \ref{Fig4} depicts the temperature dependence of MIN and F-MIN for different magnetic fields. For this purpose, we have fixed the exchange interaction $J/k_B=141 K$, $g_1=2.20$ and $g_2=2.29$  based on the experimental results\cite{Hagiwara1999}. We confirm that the correlation between  Cu and Ni magnetic ions have different variations depending on the strength of magnetic fields. At low magnetic fields, MINs are maximum at zero temperature and decrease with increase of temperature. The correlation vanishes at higher temperatures. Interestingly, we observe that the F-MIN does  exist even at  room temperature which may have wider ramifications from an experimental perspective. MINs are more robust to  thermal fluctuations and exist even at  room temperature in stark contrast to entanglement\cite{PRBMixed}.  In addition, one understands that we can sustain F-MIN even at higher temperatures compared to MIN. At higher magnetic fields, say $B=150 T$,  both MIN and F-MIN become zero  at zero temperature  and the correlation measures are unconventionally induced with the increase of  temperature which is in stark contrast to what is being observed in spin systems \cite{EntPRA}
\begin{figure*}[!ht]
\centering\includegraphics[width=0.45\linewidth]{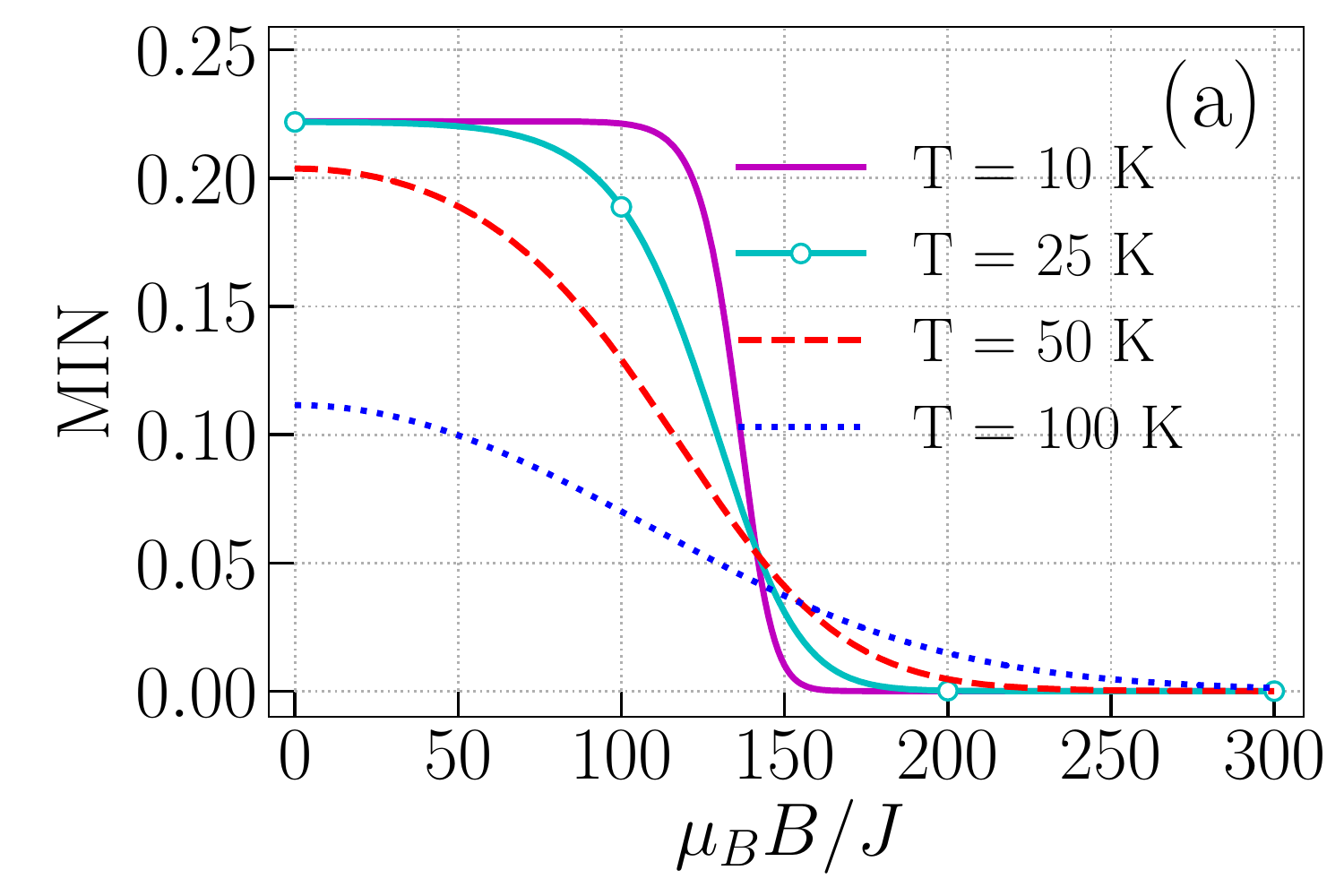}
\centering\includegraphics[width=0.45\linewidth]{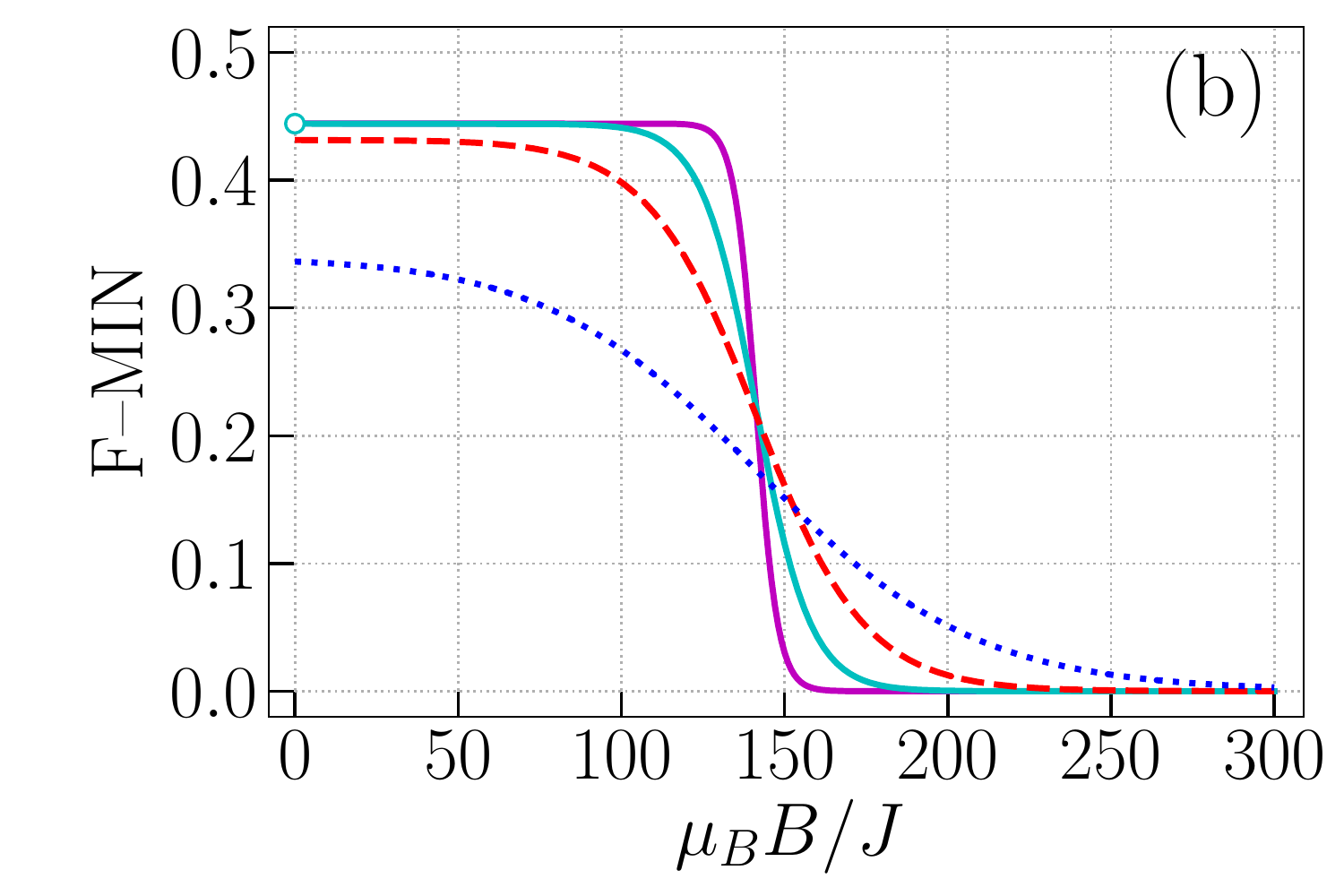}
\caption{ Thermal quantum correlations of CuNi complex quantified by MIN and F-MIN as a function of external magnetic field for  different temperatures. The fixed parameters are  $g_1 = 2.2$,  $g_2=2.29$, $J/k_B=141 K$, ~ and ~$ D/k_B = 0$.}
\label{Fig5}
\end{figure*}

\begin{figure*}[!ht]
\centering\includegraphics[width=0.45\linewidth]{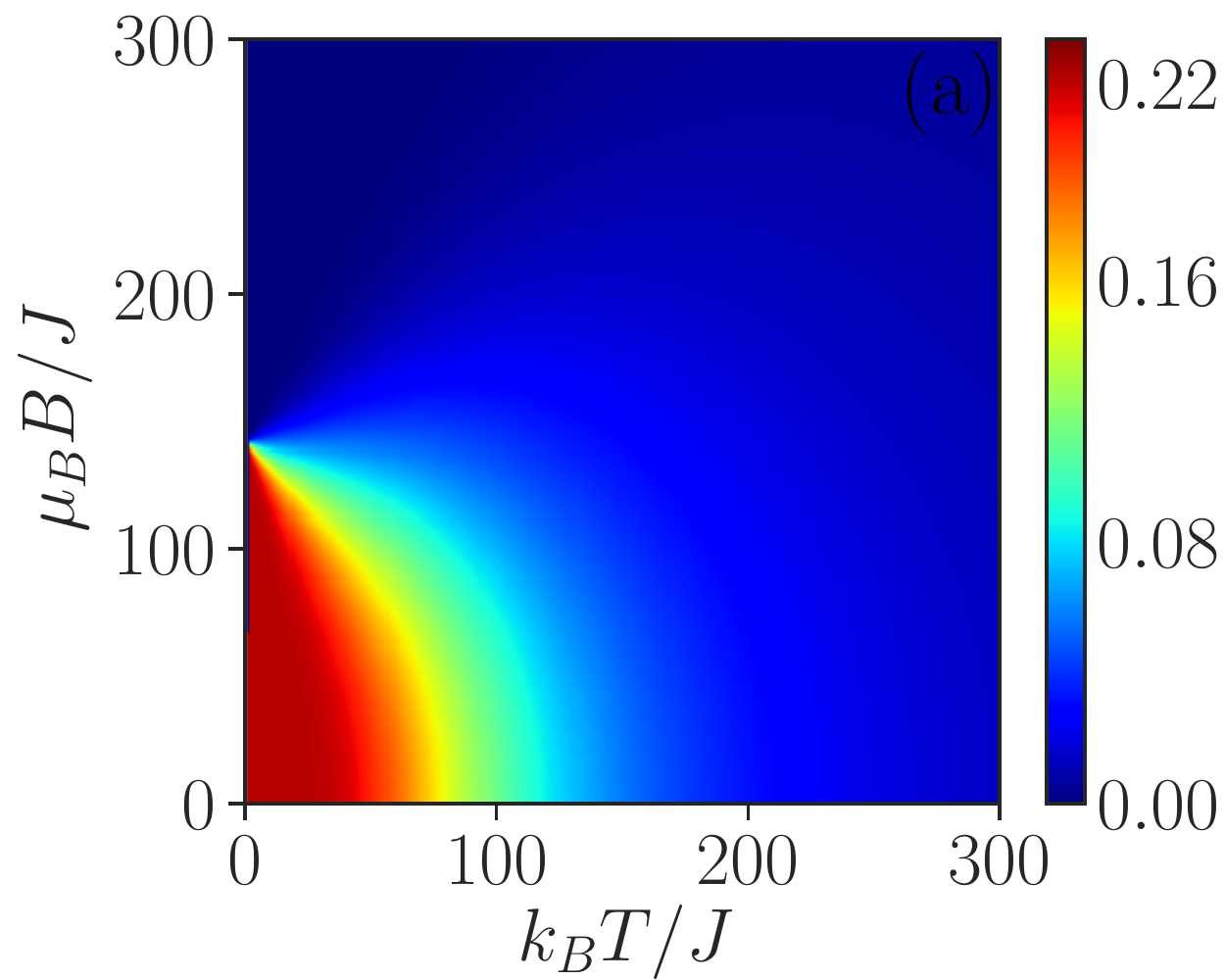}
\centering\includegraphics[width=0.45\linewidth]{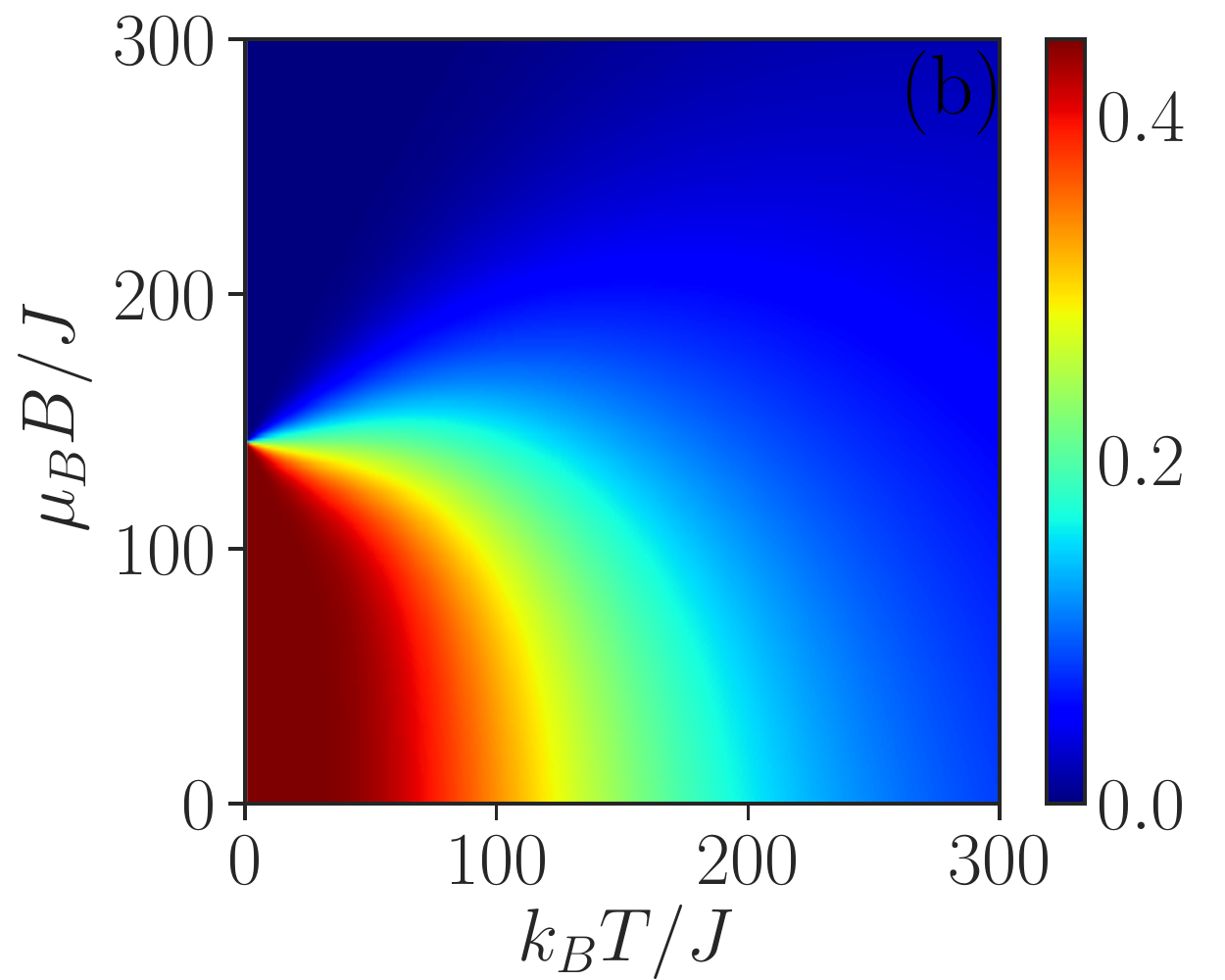}
\caption{ Thermal quantum correlations of CuNi complex quantified by MIN and F-MIN as a function of external magnetic field and temperature. The fixed parameters $g_1 =2.2$, $J/k_B = 141 K$, $g_2=2.29$, $\Delta=1$ ~and~$D/k_B=0$.}
\label{Fig6}
\end{figure*}

The correlation between the magnetic ions are also plotted as a function of magnetic field for  different temperatures for the same set of  parameters used in Fig. \ref{Fig5}. At zero magnetic field, the thermal correlation between  Cu and Ni magnetic ions is maximum and remains constant upto a critical magnetic field and  then drops to zero. In addition, we notice that increase of temperature also diminishes the correlations. In  Fig. \ref{Fig6}, we show the variation of densities as a functions of magnetic field and temperature. The quantum correlation measures exist at higher temperatures compared to   entanglement.


\section{Conclusion}
\label{concl}
To summarize, we have studied the thermal quantum correlation quantified by the measurement-induced nonlocality (MIN) in the mixed spin (1/2,1) Heisenberg dimer which have been  experimentallly realized in heterodinuclear complex. We have considered two kinds of MIN such as Hilbert-Schmidt MIN and fidelity based MIN and shown that the correlation depends on the system parameters.  It is also noticed  that the MIN can be enhanced by intrinsic parameters such as single-ion anisotropy and exchange coupling. On the other hand, the extrinsic parameters such as magnetic field and temperature decreases the quantum correlation. Under suitable parametric restrictions, we highlight that the  both the measures increase unconventionally  with the magnetic field due to Zeeman’s splitting. 

In addition, we have also investigated the quantum correlation in heterodinuclear complex CuNi which provides the experimental realization of mixed spin-(1/2, 1) Heisenberg dimer. While comparing with the entanglement, MIN can survive relatively at higher temperatures $(T=300K)$ and magnetic fields $(B=150T)$. The robustness of MIN against thermal fluctuations and magnetic field may offer huge potential  in quantum information processing through the  CuNi heterodinuclear complex.

The density matrix elements of the physical system are realizable in terms of spin observables and the fidelity is also measureable using quantum circuits \cite{subf}. Hence, we believe that our results offer more insight into the development of quantum technology with regard to information processing and its practical applications. 
\noindent

\section*{Acknowledgment}
 SB and RR wish to thank  Council of Scientific and Industrial Research (CSIR), Government of India for financial support under Grant No. 03(1456)/19/EMR-II. RM acknowledges the financial support received from Czech Technical University in Prague, Czech Republic under research grant 122-1225204D002.

\end{document}